\documentclass[11pt,draftclsnofoot,onecolumn]{IEEEtran}

\usepackage[pdftex]{graphicx}
\usepackage{subfig}
\graphicspath{{./Figures/}}
\usepackage[T1]{fontenc}
\usepackage{amsmath,amssymb,amsfonts,mathrsfs,bm}
\usepackage{amsthm}
\usepackage[font=small]{caption}
\usepackage{dsfont}
\usepackage{algorithm,algorithmic}
\usepackage{cite}
\usepackage{color}
\usepackage[normalem]{ulem}
\usepackage{array}
\newcolumntype{L}[1]{>{\raggedright\let\newline\\\arraybackslash\hspace{0pt}}m{#1}}
\newcolumntype{C}[1]{>{\centering\let\newline\\\arraybackslash\hspace{0pt}}m{#1}}
\newcolumntype{R}[1]{>{\raggedleft\let\newline\\\arraybackslash\hspace{0pt}}m{#1}}

\usepackage[utf8]{inputenc}

\usepackage[shortlabels]{enumitem}
\usepackage{breqn}

\ifx\useTheoremCounter\undefined
\newtheorem{Theorem}{Theorem}
\newtheorem{Corollary}{Corollary}
\newtheorem{Proposition}{Proposition}
\newtheorem{Lemma}{Lemma}
\else
\newtheorem{Theorem}{Theorem}

\fi

\newtheorem{Assumption}{Assumption}

\newcommand{\ofrac}[1]{{\frac{1}{#1}}}
\newcommand{\inflim}[1]{\lim\limits_{#1 \rightarrow \infty}}
\newcommand{\blderr}[1]{{\widetilde{\Bs{#1}}}}

\newcommand{\Norm}[1]{\left\| {#1} \right\| }
\newcommand{\Diag}[1]{\operatorname{diag}\left\{{#1}\right\}}
\newcommand{\Col}[1]{\operatorname{col}\left\{{#1}\right\}}
\newcommand{\mc}[1]{\mathcal{{#1}}}

\newcommand{\Bs}[1]{\boldsymbol{#1}}
\newcommand{\bpsi}{\bm{\psi}}
\newcommand{\bphi}{\bm{\phi}}
\newcommand{\bw}{\bm{w}}
\newcommand{\bu}{\bm{u}}
\newcommand{\bv}{\bm{v}}
\newcommand{\bd}{\bm{d}}
\newcommand{\bx}{\bm{x}}
\newcommand{\bh}{\bm{h}}
\newcommand{\bW}{\bm{W}}
\newcommand{\btau}{\bm{\tau}}
\newcommand{\bOmega}{\bm{\Omega}}

\newcommand{\lk}{{\ell k}}
\newcommand{\E}{\mathbb{E}}
\newcommand{\CalA}{\mathcal{A}}
\newcommand{\CalB}{\mathcal{B}}
\newcommand{\CalC}{\mathcal{C}}

\newcommand{\CalE}{\mathcal{E}}
\newcommand{\CalF}{\mathcal{F}}
\newcommand{\CalG}{\mathcal{G}}

\newcommand{\CalM}{\mathcal{M}}
\newcommand{\CalN}{\mathcal{N}}

\newcommand{\CalR}{\mathcal{R}}
\newcommand{\CalS}{\mathcal{S}}

\newcommand{\CalX}{\mathcal{X}}
\newcommand{\CalY}{\mathcal{Y}}

\newcommand{\BCalB}{\Bs{\mathcal{B}}}

\newcommand{\BCalR}{\Bs{\mathcal{R}}}

\newcommand{\NI}{\CalN_{I,k}}
\newcommand{\NC}{\CalN_{C,k}}
\newcommand{\NCell}{\CalN_{C,\ell}}
\newcommand{\Tr}{\operatorname{Tr}}
\newcommand{\Vc}{\operatorname{vec}}
\newcommand{\cO}{\mathcal{O}}

\definecolor{mypurple}{rgb}{0.49, 0.18, 0.56}
\definecolor{myorange}{rgb}{0.99, 0.35, 0.11}

\begin{document}
%
\title{A Multitask Diffusion Strategy with Optimized Inter-Cluster Cooperation}
%
%
%

\author{Yuan~Wang,~\IEEEmembership{Student~Member,~IEEE,}
        Wee~Peng~Tay,~\IEEEmembership{Senior~Member,~IEEE,}
        and~Wuhua~Hu,~\IEEEmembership{Member,~IEEE,}
        \thanks{This work was supported in part by the Singapore Ministry of Education Academic Research Fund Tier 2 grants MOE2013-T2-2-006 and MOE2014-T2-1-028, and the Delta-NTU Corporate Lab through the National Research Foundation (NRF) corporate lab@university scheme.} 
        \thanks{Y. Wang and W. P. Tay are with the School of Electrical and Electronic Engineering, Nanyang Technological University, Singapore. Emails: ywang037@e.ntu.edu.sg, wptay@ntu.edu.sg.} 
        \thanks{W. Hu was with the School of Electrical and Electronic Engineering, Nanyang Technological University, Singapore, and now is with the Signal Processing Department, Institute for Infocomm Research, Agency for Science, Technology and Research (A*STAR), Singapore. E-mail: huwh@i2r.a-star.edu.sg.}
        }
\maketitle

\begin{abstract}
We consider a multitask estimation problem where nodes in a network are divided into several connected clusters, with each cluster performing a least-mean-squares estimation of a different random parameter vector. Inspired by the adapt-then-combine diffusion strategy, we propose a multitask diffusion strategy whose mean stability can be ensured whenever individual nodes are stable in the mean, regardless of the inter-cluster cooperation weights. In addition, the proposed strategy is able to achieve an asymptotically unbiased estimation, when the parameters have same mean. We also develop an inter-cluster cooperation weights selection scheme that allows each node in the network to locally optimize its inter-cluster cooperation weights. Numerical results demonstrate that our approach leads to a lower average steady-state network mean-square deviation, compared with using weights selected by various other commonly adopted methods in the literature.
\end{abstract}


%

\section{Introduction}\label{sec:introduction}
\IEEEPARstart{D}{istributed} learning and inference over multi-agent networks has been applied to a broad range of applications like distributed localization, target tracking, spectrum sensing and adaptive learning \cite{XuQuiLen:J15,ZhaTayLi:J14,LM15TSP,SYTU13TSP,HoTayQue:J15,CS11TSP}. Although centralized methods exploit more information aggregated from all the nodes (or agents) in the network and may result in better performance, they suffer from some crucial limitations. In these methods, nodes in the network send information to a central fusion center for processing, which leads to high communication costs. More importantly, such methods are vulnerable to the failure of the fusion center which may cause the collapse of the entire system. By contrast, in distributed adaptive networks, the fusion center is no longer needed. Agents in the network process information locally, and exchange summary statistics with their neighbors, so as to achieve better inference performance compared with the case where each agent performs inference relying only on its own local data.  

Various distributed solutions have been proposed in the literature, including incremental strategies \cite{CGL07T,DB1997,NB14T}, and consensus strategies \cite{BYD06,OLF07T,NDC10T}. Compared with the previous two categories, diffusion adaptation strategies \cite{CAT10TSP,AHS12TR,XCZ12TSP,ONG13STSP,WH15TSP,ARB14JANTSP,SMO14TSP,ZJTOWFIC14TSP,HSLEE15TSP,YWANG15ICICS,JIECHEN16ICASSP} are advantageous under mild technical conditions \cite{SYT12TSP,SYT13SPM,AHS14Proc}. Specifically, diffusion strategies adopt constant step-sizes that endow the network with continuous adaptation and learning ability, whereas the time-dependent or decaying step-sizes are usually used in the single time-scale consensus-based counterparts to achieve consensus among nodes at the expense of adaptability to changing parameter values. For tasks such as adaptive learning from streaming data under constant step-sizes, both mean and mean-square stability of diffusion networks is insensitive to the choice of combination weights used for scaling the information received at each node, and diffusion networks can achieve better steady-state network mean-square deviation (MSD) than single time-scale consensus networks with constant step-sizes \cite{SYT12TSP}. It is also shown in \cite{SYT12TSP,SYT13SPM,AHS14Proc} that consensus networks can become unstable even if all the individual nodes are stable and able to solve the estimation task on their own. 

Most of the abovementioned works on diffusion strategies assume that all the nodes in the network share a single common parameter of interest. However, in practical multi-agent networks, different nodes or clusters of nodes may have different parameters of interest \cite{NB14T,JPLATA15TSP,NB15ICASSP,JPLATA16ICASSP,XCZ15TSP,JC15TSP}. In some applications, the parameters of interest at each node may be different or may overlap. The reference \cite{NB14T} considers an estimation problem where each node estimates three categories of parameters: a global parameter, parameters of local interest, and parameters of common interest to a subset of nodes. A similar nodes-specific distributed estimation problem is solved by a diffusion-based strategy in \cite{JPLATA15TSP}, and is studied from a coalition game theoretic perspective in \cite{NB15ICASSP}. In the work \cite{JPLATA16ICASSP}, the relationship between nodes-specific estimation tasks are assumed to be unknown, and each node infers which parameters their neighbors are estimating in order to adjust their range of cooperation. In the aforementioned works, each node only cooperates with other nodes that have common parameters of interest. In some other applications, nodes may have different parameters of interest, which may be correlated with each other. For example, each cluster of nodes may be tasked to track a particular target. In \cite{JC15TSP}, nodes are not aware of whether their neighbors perform the same task. To reduce the resulting bias, an on-line combination weight adjustment scheme was proposed to cluster the nodes. By allowing inter-cluster cooperation, clusters can achieve better performance than by operating independently of each other. To address this \emph{multitask}-oriented distributed estimation problem, \cite{JC14C,JC14TSP} propose a multitask diffusion adaptation strategy that inherits the basic structure of the adapt-then-combine (ATC) strategy \cite{AHS12TR}, and allows cooperation among clusters by adding an additional $\ell_2-$norm regularization to the ATC objective function. The reference \cite{RNASSIF15ICASSP} uses $\ell_1-$norm regularization to promote sparsity in the difference between cluster parameters. An asynchronous network model for the multitask diffusion strategy is considered in \cite{RNASSIF16TSP}, where nodes may not update and exchange estimates synchronously. In the \cite{JC14TSP,RNASSIF15ICASSP,RNASSIF16TSP}, the network MSD performance is controlled by the regularization weight and the cooperation weights between clusters. However, how to select these weights was left as an open problem. A heuristic adaptive method for selecting the inter-cluster cooperation weights was proposed in \cite{SMONAJEMI15MLSP}. How to choose the regularization and cooperation weights in a rigorous optimal way has not been adequately addressed. In particular, we note that certain choices of these weights may lead to worse-off performance of some of the clusters compared to the non-cooperation case (see example in Section \ref{sec:Sim}). 

In this paper, we consider a distributed multitask least-mean-square (LMS) estimation problem for random parameters. Our main contributions are as follows:
\begin{itemize}
\item We propose a multitask diffusion strategy that incorporates an intermediate inter-cluster diffusion combination step between the local update and the conventional intra-cluster diffusion combination steps. This strategy combines the regularization weight and the inter-cluster cooperation weights in \cite{JC14TSP,RNASSIF15ICASSP,RNASSIF16TSP} so that the network performance is now influenced by the inter-cluster cooperation weights only, and tuning the regularization weight is no longer needed. 
\item We show that the mean stability of our proposed strategy can be achieved independent of the choice of the inter-cluster cooperation weights, in contrast to \cite{JC14C,JC14TSP} whose stability depends on these weights. This resembles the advantage of the ATC strategy versus the consensus strategy, where the mean stability of ATC is not affected by the choice of combination weights \cite{SYT12TSP,SYT13SPM,AHS14Proc}. In fact, the conditions to guarantee the mean stability of our proposed multitask diffusion strategy is the same as that of the conventional ATC diffusion strategies, which implies that as long as each node or each cluster is stable in the mean, then incorporating inter-cluster cooperation preserves the mean stability of the network.
\item We propose a centralized and a distributed inter-cluster cooperation weight optimization scheme to optimize the steady-state network MSD performance. An adaptive online approach is also proposed when certain statistics used in the optimization methods are not known a priori. 
\end{itemize}
This work is a comprehensive extension of \cite{YWANG16SSP} which presents only the centralized and distributed optimization schemes with limited performance analysis and experiment results. 

The rest of this paper is organized as follows. In Section \ref{sec:Pre}, we introduce the network model and problem formulation, discuss some prior works, and describe our proposed multitask strategy. In Section \ref{sec:analysis} we examine the performance of the proposed strategy via mean and mean-square error behavior analysis. In Section \ref{sec:Weights}, we present optimization schemes to select the inter-cluster cooperation weights and an adaptive implementation. Finally, numerical results and the conclusion follow in Sections \ref{sec:Sim} and \ref{sec:Conclusion}, respectively.

\emph{Notation.} 
Throughout this paper, we use boldface characters for random variables, and plain characters for realizations of the corresponding random variables as well as deterministic quantities. Besides, we use upper-case characters for matrices and lower-case ones for vectors and scalars. For ease of comparison, we also adopt similar notations used in \cite{AHS12TR,JC14TSP}. The notation $\mathds{1}_N$ represents an $N\times 1$ vector with all entries being one, and $I_N$ is an $N\times N$ identity matrix. The vector $0_M$ is an $M\times1$ vector with all entries being zero. The matrix $A^T$ is the transpose of the matrix $A$, $\lambda_{\max}(A)$ is the largest eigenvalue of the matrix $A$, and $\rho(A)$ is the spectral radius of $A$. The operation $A\otimes B$ denotes the Kronecker product of the two matrices $A$ and $B$. The notation $\Norm{\cdot}$ is the Euclidean norm, $\Norm{\cdot}_{b,\infty}$ denotes the block maximum norm \cite{AHS12TR}, while $\Norm{x}^2_\Sigma \triangleq x^T\Sigma x$ for any column vector $x$ and non-negative definite matrix $\Sigma$. We use $(x_i)_{i=1}^n$ to represent the sequence of vectors or scalars $x_1,\ldots,x_n$, while $(A_i)_{i=1}^n$ represents the sequence of matrices $A_1,\ldots,A_n$. We also use $[x_i]_{i=1}^n$ to denote the matrix $[x_1,\ldots,x_n]$. We use $\Diag{\cdot}$ to denote a matrix whose main diagonal is given by its arguments, and $\Col{\cdot}$ to denote a column vector formed by its arguments. The notation $\Vc(\cdot)$ represents a column vector consisting of the columns of its matrix argument stacked on top of each other. If $\sigma = \Vc(\Sigma)$, we let $\Norm{\cdot}_\sigma = \Norm{\cdot}_\Sigma$, and use either notations interchangeably. In addition, we use $x \geq y$ to represent entry-wise inequality between vectors $x$ and $y$, and $\big|\cdot\big|$ is the number of the elements of its set argument.
\begin{figure}[!t]
\centering
\includegraphics[width=2.8in]{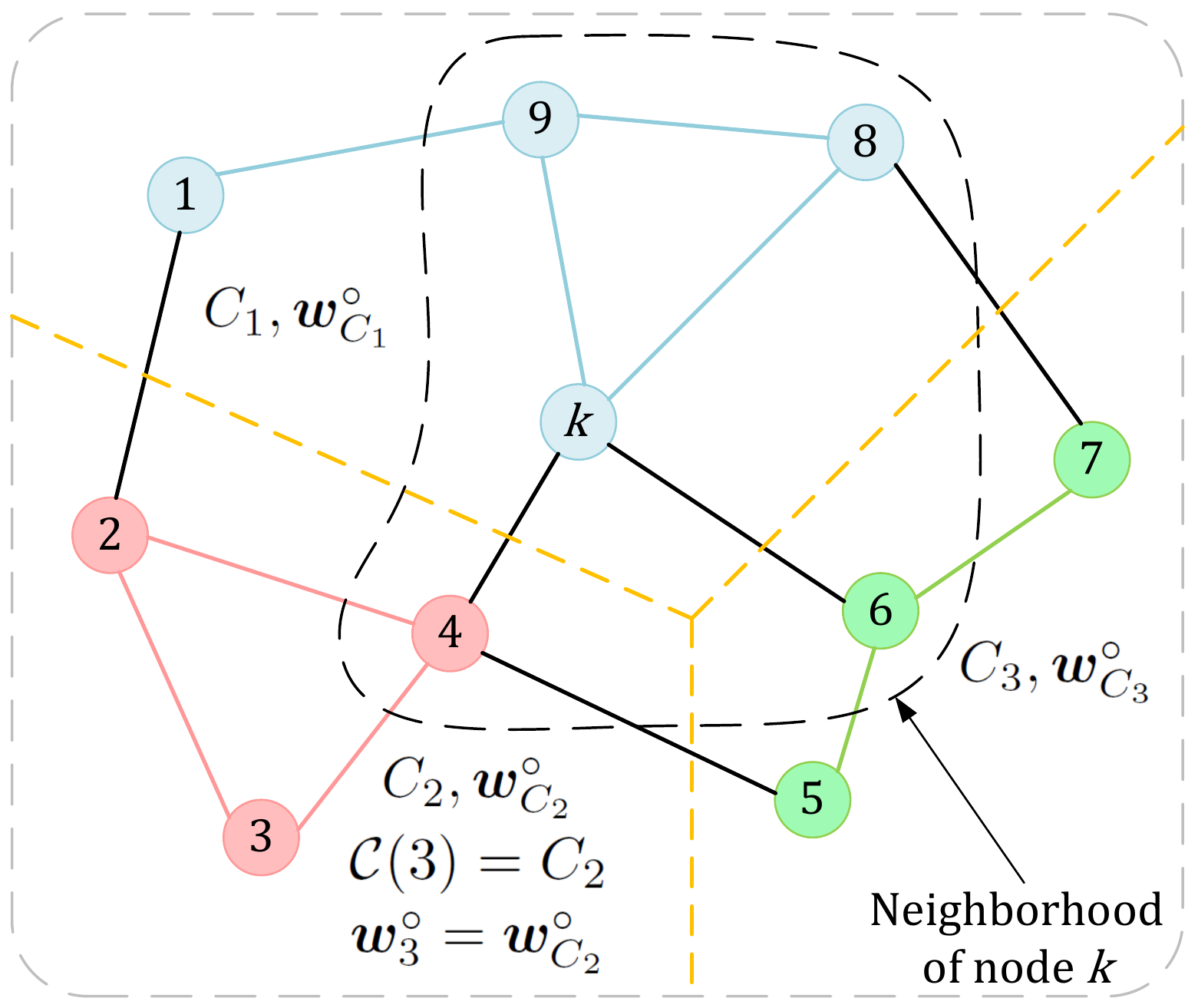}
\caption{An example of a clustered multitask network with three clusters depicted in red, blue, and green circles, respectively.}
\label{fig:NetEg}
\end{figure}

\section{Problem Formulation and Multitask Diffusion}\label{sec:Pre}
In this section, we first present our network and data model, give a brief overview of the multitask diffusion strategy proposed by \cite{JC14TSP}, and finally present our proposed multitask diffusion strategy.

\subsection{Network and Data Model}\label{ssec:Model}
We consider a distributed adaptive network with $N$ nodes (see example depicted in Fig.~\ref{fig:NetEg}). The network is represented by an undirected graph where the vertices denote the nodes, any two nodes are said to be connected if there is an edge between them. The neighborhood of any particular node $k$ is denoted by $\CalN_k$, which consists of all the nodes that are connected to node $k$, and node $k$ itself. Since the network is assumed to be undirected, if node $k$ is a neighbor of node $\ell$ then node $\ell$ is also a neighbor of node $k$. Without loss of generality, we assume that the network is connected.

In the context of multitask learning over clustered networks, nodes are grouped into $P$ clusters, each belonging to a unique cluster (as illustrated in Fig.~\ref{fig:NetEg}). Let the clusters be indexed by $\mathcal{P}\triangleq \{1,2,\ldots,P\}$. Each cluster $C_p$, $p\in \mathcal{P}$, aims to estimate an unknown parameter vector $\Bs{w}_{C_p}^\circ\in\mathbb{R}^{M\times 1}$. For a node $k$ belonging to a cluster $C_p$ for some $p\in\mathcal{P}$, we use $\CalC(k)$ to refer to this specific cluster. The intra-cluster neighborhood $\NC$ consists of all the neighboring nodes that belong to the same cluster as node $k$, which includes node $k$ itself. The inter-cluster neighborhood $\NI$ consists of neighboring nodes of $k$ that belong to clusters different from $\CalC(k)$. We have
\begin{align}
\NC&=\CalN_k\cap\CalC(k), \\
\NI&=\CalN_k\backslash\CalC(k),
\end{align} 
and
\begin{align}
\NC\cup\NI&=\CalN_k, \\
\NC\cap\NI&=\emptyset.
\end{align}  

In this paper, we suppose the parameters $\Bs{w}^\circ_{C_p}$, $p\in\mathcal{P}$, to be \emph{random} and correlated with each other. This is in contrast to most of the literature \cite{AHS12TR,JC14TSP,SYT13SPM}, which assume that parameters of interest are deterministic. Thus, our objective is then to characterize the expected network estimation performance, where the expectation is not only taken over the random realizations of the data and noise, but also the random parameters. Although \cite{XZHAO12TSP} also assumes that parameters are random and follow a random walk model, the network considered is of \emph{single}-task, namely every node in the network estimates the same parameter of interest.

For a node $k \in C_p$, let $\Bs{w}_k^\circ=\Bs{w}_{C_p}^\circ$. At each time instant $i\ge1$, each node $k$ observes a scalar random variable $\bd_{k,i}\in\mathbb{R}$, and a random vector $\Bs{u}_{k,i}\in\mathbb{R}^{M\times 1}$.\footnote{Throughout this paper, all vectors are column vectors.} The observations of node $k$ are related to $\Bs{w}_k^\circ$ via the linear regression model:
\begin{align}
\bd_{k,i} = \Bs{u}_{k,i}^T\Bs{w}_k^\circ + \Bs{v}_{k,i}, \label{eq:datamodel}
\end{align}
where $\Bs{v}_{k,i}$ is a zero-mean random noise. Note that although $\Bs{w}_k^\circ$ is random, for our steady-state analysis, it does not change over the time instants $i$. We make the following assumptions.

\begin{Assumption}\label{asmp:regressor}
The regression process $\{\bu_{k,i}\}$ is zero-mean, spatially independent and temporally white. The regressor $\bu_{k,i}$ has positive definite covariance matrix $R_{u,k}=\E\Bs{u}_{k,i}\Bs{u}_{k,i}^T$.  
\end{Assumption}

\begin{Assumption}\label{asmp:noise}
The noise process $\{\Bs{v}_{k,i}\}$ is spatially independent and temporally white. The noise $\Bs{v}_{k,i}$ has variance $\sigma^2_{v,k}$, and is assumed to be independent of the regressors $\Bs{u}_{\ell,j}$ for all $\{k, \ell, i, j\}$.
\end{Assumption}

\begin{Assumption}\label{asmp:para}
The parameters of interest $\{\Bs{w}^\circ_k\}^N_{k=1}$ are independent of $\{\bu_{\ell,i},\bv_{\ell,i}\}$ for all $\{k, \ell, i\}$. 
\end{Assumption}
In this paper, we assume that all the cluster parameters take values from the same compact space $\mathcal{W}$, and we do not have any prior information about their distributions. Therefore, we assume that the cluster parameters are uniformly distributed over $\mathcal{W}$, and thus have the same mean. An example is when clusters are assigned to track different targets, which may appear randomly within a specified region. The assumption that cluster parameters have the same mean can also model the case where we know a priori the mean of each cluster parameter so that this can be subtracted out from the data model \eqref{eq:datamodel}. For example, a cluster may be assigned to monitor a specific sub-region of interest, and a target may appear within that sub-region with some known distribution. Therefore, we make the following assumption throughout this paper.
\begin{Assumption}\label{asmp:mean}
Let $\Bs{w}^\circ=\Col{(\Bs{w}^\circ_k)^N_{k=1}}$. We assume that all the random parameters $\{\Bs{w}^\circ_k\}^N_{k=1}$ have the same mean $\overline{w}$, i.e., $\E\Bs{w}^\circ= \mathds{1}_N\otimes\overline{w}$.
\end{Assumption}

Let the second-order moment of $\Bs{w}^\circ$ be
\begin{align}
\CalR_w = \E\Bs{w}^\circ{\Bs{w}^\circ}^T . \label{eq:calRw}
\end{align}
Note that the parameters $\{\Bs{w}^\circ_{C_p}\}_{p\in \mathcal{P}}$ may be correlated in general. For example in the aforementioned target tracking problem, if each cluster is tracking a different target and the targets are moving in tandem with each other, then the cluster parameters are correlated with each other. Likewise, when the temperature of an area needs to be estimated, the temperature of each position can be different but correlated with each other. It is thus beneficial to seek cooperation between different clusters by allowing nodes in different clusters to exchange information with each other. We seek an inter-cluster cooperation scheme that minimizes the network MSD \emph{on average} over all realizations of the cluster parameters. As will become obvious in the sequel, the optimal inter-cluster cooperation depends on $\CalR_w$. We will first assume that the correlation matrix $\CalR_w$ is known a priori, and then provide an adaptive implementation that estimates $\CalR_w$ in each time step.

\subsection{Multitask Diffusion Strategy with Regularization}\label{ssec:MDLMS}
Conditioned on $\Bs{w}_{\CalC(k)}=w_{\CalC(k)}$ for all $k$,  in \cite{JC14TSP}, each node $k$ in cluster $\CalC(k)$ is associated with the following local cost function:
\begin{dmath}
J_{k}\left(w_{\CalC(k)}\right) = \E \left[\left|\bd_{k,i} - \bu_{k,i}^T \Bs{w}_{\CalC(k)} \right|^2 \middle| {\Bs{w}_{\CalC(k)}=w_{\CalC(k)}} \right] + \eta \sum_{\ell\in\NI}  \rho_{\lk} \left\| w_{\CalC(k)} - w_{\CalC(\ell)} \right\|^2 , \label{eq:cost}
\end{dmath}
where $\eta$ and $\{ \rho_{\lk}\}$ are non-negative regularization weights. To minimize the sum of the local cost functions in \eqref{eq:cost}, the work \cite{JC14TSP} proposes a multitask diffusion LMS strategy (MDLMS) which takes the following form: 
\begin{align}
\begin{cases}
\bpsi_{k,i} &= \bw_{k,i-1} + \mu_k \bu_{k,i} ( \bd_{k,i} - \bu_{k,i}^T \bw_{k,i-1})  \\
& \qquad + \mu_k  \sum \limits_{\ell \in \NI } \eta \rho_{\lk} \left( \bw_{\ell,i-1} - \bw_{k,i-1} \right) \\ 
\bw_{k,i} &= \sum \limits_{\ell \in \NC} a_{\ell k} \bpsi_{\ell,i}
\end{cases} \label{eq:CJAlgo}
\end{align}
where $\mu_k$ is a constant step-size and the scalar intra-cluster combination weights $\{a_{\ell k}\}$ satisfy
\begin{align}
{a_{\ell k}} \ge 0, & {\sum \limits_{\ell \in \NC } a_{\ell k} = 1}, \; {a_{\ell k} = 0}, \;  \text{if} \; {\ell \notin \NC}. \label{eq:alk} 
\end{align}
Likewise, the scalar inter-cluster cooperation weights $\{ \rho_{\lk}\}$ satisfy 
\begin{align}
{ \rho_{\lk}} \ge 0, & {\sum \limits_{\ell \in \NI}} \; { \rho_{\lk} = 1}, \; { \rho_{\lk} = 0}, \; \text{if} \; \ell \notin \NI{,}  \label{eq:rho}
\end{align}
Compared with the local update equation of the ATC strategy \cite{SYT13SPM}, the additional third term on the right-hand side (R.H.S.) of the first update equation of \eqref{eq:CJAlgo} is introduced to extend the cooperation to between clusters by taking the estimates $ \bw_{\ell,i-1} $ from neighboring nodes in different clusters. Thus, more information is utilized during distributed adaptation, and the network estimation performance could be improved by proper selection of $\eta$ and $\{ \rho_{\lk}\}$. In \cite{JC14TSP}, the inter-cluster cooperation weights are chosen to be
\begin{align}\label{eq:averagerule}
 \rho_{\lk}  =
\begin{cases}
{\left| \NI \right|}^{-1}, \; & \text{if} \; \ell \in \NI,  \\
0, & \text{otherwise,}
\end{cases} 
\end{align}
for ease of implementation. No method has been proposed in \cite{JC14TSP} to find the best cooperation weights. We will refer to \eqref{eq:averagerule} as the \emph{inter-cluster averaging rule} in the sequel.

\subsection{Proposed Multitask Diffusion Strategy with Adaptation before Inter-cluster Cooperation}\label{sec:Proposed}
In this subsection, we propose a multitask diffusion strategy that performs adaptation before inter-cluster cooperation. We show in Section \ref{sec:analysis} that our proposed strategy can achieve asymptotically unbiased estimation and mean stability irrespective of the inter-cluster cooperation weights under the assumptions in Section \ref{ssec:Model}. 

Conditioned on $\Bs{w}_{\CalC(k)}=w_{\CalC(k)}$ for all $k$, we consider the same local cost function in \eqref{eq:cost}. Proceeding in a similar fashion as \eqref{eq:CJAlgo}, we can rewrite the first equation of \eqref{eq:CJAlgo} as:
\begin{align}
\begin{cases}
\bpsi_{k,i} &= \bw_{k,i-1} + \mu_k \bu_{k,i} ( \bd_{k,i} - \bu_{k,i}^T \bw_{k,i-1}),  \\
\bphi_{k,i} &= \bpsi_{k,i} - \mu_k \sum \limits_{\ell \in \NI} \eta  \rho_{\lk} (\bw_{k,i-1} - \bw_{\ell,i-1}).
\end{cases} \label{eq:replace}
\end{align}
Since the value $\bpsi_{k,i}$ is the updated local estimate and thus a better estimate for $\bw^\circ_k$ than $\bw_{k,i-1}$ \cite{AHS12TR} , we use $\bpsi_{k,i}$ and $\bpsi_{\ell,i}$ to replace $\bw_{k,i-1}$ and $\bw_{\ell,i-1}$, respectively in the second equation of \eqref{eq:replace}. Let
\begin{align}
g_{\ell k} &= \mu_k \eta \rho_{\lk}, \text{ if } \ell \in \NI, \nonumber \\
g_{kk} &= 1 - \mu_k \sum \limits_{\ell \in \NI} \eta \rho_{\lk} {.} \label{eq:gamma_definition}
\end{align}
These eliminate the redundant degree of freedom offered by $\eta$ which complicates the optimization of the inter-cluster cooperation weights $\{ \rho_{\lk}\}$ in \eqref{eq:rho}. We obtain the following Multitask Adapt, Inter-cluster cooperation, and then Combine (MAIC) diffusion strategy:
\begin{align}
\begin{cases}
\bpsi_{k,i} &= \bw_{k,i-1} + \mu_k \bu_{k,i} ( \bd_{k,i} - \bu_{k,i}^T \bw_{k,i-1})  \\
\bphi_{k,i} &= \sum \limits_{\ell \in \NI^+} g_{\ell k}\bpsi_{\ell,i}  \\
\bw_{k,i} &= \sum \limits_{\ell \in \NC} a_{\ell k}\bphi_{\ell,i} 
\end{cases} \label{eq:newalgo}
\end{align}
where $\NI^+ = \NI\cup\{k\}$ and the inter-cluster weights $\{g_{\ell k}\}$ satisfy
\begin{align}
g_{\ell k}\ge0,\; {\sum \limits_{\ell \in \NI^+}} \; {g_{\ell k} = 1}, \; {g_{\ell k} = 0}, \; \text{if} \; \ell \notin \NI^+ {.}  \label{eq:gamma_condition}
\end{align}
Let $G$ be a matrix with the $(\ell,k)$-th entry being $g_{\ell k}$. Then, from \eqref{eq:gamma_definition}, the matrix $G$ is left-stochastic, i.e., $G^T \mathds{1}_N = \mathds{1}_N$. The proposed diffusion strategy \eqref{eq:newalgo} is in the same spirit as the ATC diffusion strategy, which performs adaptation at each node before combining information from neighboring nodes. Similarly, MAIC first performs adaptation, before cooperatively integrating information from neighboring nodes not within its cluster. Finally, it performs combination of information from neighboring nodes within its own cluster.

\section{Performance Analysis}\label{sec:analysis}
In this section, we study the error behavior of the proposed MAIC diffusion strategy, in terms of steady-state mean and mean-square performance. We derive sufficient conditions for mean stability and mean-square stability, and the steady-state network MSD. For ease of reference, we summarize the commonly used symbols in Table \ref{tbl:notations}.

\begin{table}[!h]
\renewcommand{\arraystretch}{1.5}
\caption{Definitions of commonly used symbols.}
\label{tbl:notations}
\centering
\begin{tabular}{ l|c }
  \hline
  Symbol & Equation\\ \hline
  $\CalA = A \otimes I_M$ & \eqref{eq:calA} \\
  $\CalG = G \otimes I_M$ & \eqref{eq:calG} \\
  $\BCalR_{u,i} = \Diag{(\bu_{k,i}\bu_{k,i}^T)^N_{k=1}}$ & \eqref{eq:calRui} \\
  $\CalR_u = \E\BCalR_{u,i}=\Diag{(R_{u,k})^N_{k=1}}$  & \eqref{eq:calRu} \\ 
  $\CalM = \Diag{(\mu_k I_M)^N_{k=1}}$ & \eqref{eq:calM} \\ 
  $\CalS = \Diag{(\sigma^2_{v,k} R_{u,k})_{k=1}^N}$ & \eqref{eq:matrixS} \\ 
  $\BCalB_i = \CalA^T \CalG^T (I_{MN} - \CalM \BCalR_{u,i})$ & \eqref{eq:matrixBi} \\ 
  $\CalB = \E\BCalB_i = \CalA^T \CalG^T (I_{MN} - \CalM \CalR_u)$ &\eqref{eq:matrixB} \\
  $\CalE = \E\left[ \BCalB_i^T \otimes \BCalB_i^T \right]$ & \eqref{eq:matrixE} \\ 
  $\CalF = \CalB^T \otimes \CalB^T$ & \eqref{eq:matrixF} \\
  $\CalR_w = \E\Bs{w}^\circ{\Bs{w}^\circ}^T$ & \eqref{eq:calRw} \\
  $R_{w,\lk} = \E\Bs{w}^\circ_\ell {\Bs{w}^\circ_k}^T$ & \eqref{eq:Rwlk} \\
  $\Omega_{a,k}=\Diag{\left(\mu^2\sigma^2_{v,\ell}\Tr(R_{u,\ell})\right)_{\ell\in\NI^+}}$ & \eqref{eq:Omega_ak} \\
  $\Omega_{b,k} = \E\big[\Bs{W}_{k}^T \Bs{W}_{k}\big]$ & \eqref{eq:Omega_bk}\\
  $\Bs{W}_k=[\Bs{w}^\circ_\ell]_{\ell\in\NI^+}$ & \eqref{eq:Wk} \\
  $n_k = \big|\NI^+ \big|$ & \eqref{eq:nk} \\
  \hline
\end{tabular}
\end{table}

\subsection{Network Error Recursions}\label{ssec:errorrecursion}
We first derive the recursive equations for various error vectors required for our stability analysis. The error vectors for each time instant $i$ at each node $k$ are defined as
\begin{align}
\widetilde{\Bs{\psi}}_{k,i} &= \Bs{w}^\circ_k - \Bs{\psi}_{k,i},  \nonumber\\
\widetilde{\Bs{\phi}}_{k,i} &= \Bs{w}^\circ_k - \Bs{\phi}_{k,i},  \nonumber\\
\widetilde{\Bs{w}}_{k,i} &= \Bs{w}^\circ_k - \Bs{w}_{k,i}. 
\end{align}
We collect the iterates $\widetilde{\Bs{\psi}}_{k,i}$,  $\widetilde{\Bs{\phi}}_{k,i}$, $\widetilde{\Bs{w}}_{k,i}$ across all nodes as: 
\begin{align}
\widetilde{\Bs{\psi}}_{i} &= \Col{(\widetilde{\Bs{\psi}}_{k,i})_{k=1}^N}, \nonumber \\
\widetilde{\Bs{\phi}}_{i} &= \Col{(\widetilde{\Bs{\phi}}_{k,i})_{k=1}^N}, \nonumber\\
\widetilde{\Bs{w}}_{i} &= \Col{(\widetilde{\Bs{w}}_{k,i})_{k=1}^N} {.} \label{eq:errorvec}
\end{align}

Subtracting both sides of the first equation of \eqref{eq:newalgo} from $\Bs{w}^\circ_k$, and applying the data model \eqref{eq:datamodel}, we obtain the following error recursion:
\begin{align}
\widetilde{\Bs{\psi}}_{k,i} = \left(I_M-\mu_k\bu_{k,i}\bu_{k,i}^T \right) \widetilde{\Bs{w}}_{k,i-1} - \mu_k \bu_{k,i}\bv_{k,i} {.} \label{eq:errorpsi}
\end{align}
Note that the second equation of \eqref{eq:newalgo} can be expressed as,
\begin{align}
\Bs{\phi}_{k,i} = g_{kk} \Bs{\psi}_{k,i} + \sum\limits_{\ell\in\NI}g_{\ell k} \Bs{\psi}_{\ell,i}  {.} 
\end{align}
Similarly, subtracting both sides of the above equation from $\Bs{w}^\circ_k$ leads to
\begin{dmath}
\widetilde{\Bs{\phi}}_{k,i} = g_{kk} \left( \Bs{w}^\circ_k - \Bs{\psi}_{k,i} \right) + \sum\limits_{\ell\in\NI}g_{\ell k} \left( \Bs{w}^\circ_\ell - \Bs{\psi}_{\ell,i} + \Bs{w}^\circ_k - \Bs{w}^\circ_\ell \right) 
=\sum\limits_{\ell\in\NI^+} g_{\ell k} \widetilde{\Bs{\psi}}_{\ell,i} + \sum\limits_{\ell\in\NI^+} g_{\ell k} \left( \Bs{w}^\circ_k - \Bs{w}^\circ_\ell \right) 
\end{dmath} 
Relating the above equation with the second line of \eqref{eq:errorvec} gives
\begin{align}
\widetilde{\Bs{\phi}}_i = \CalG^T\widetilde{\Bs{\psi}}_i + \left(I_{MN} - \CalG^T\right)\Bs{w}^\circ,  \label{eq:errorphi}
\end{align} 
where
\begin{align}
\CalG = G \otimes I_M . \label{eq:calG}
\end{align}
Then, subtracting the third equation of \eqref{eq:newalgo} from $\Bs{w}^\circ_k$ and using \eqref{eq:errorvec} we have
\begin{align}
\widetilde{\Bs{w}}_i = \CalA^T \widetilde{\Bs{\phi}}_i, \label{eq:errorw} 
\end{align}
where
\begin{align}
\CalA = A \otimes I_M . \label{eq:calA}
\end{align} 
Now, substituting \eqref{eq:errorphi} into the equation \eqref{eq:errorw} yields
\begin{align}
\widetilde{\Bs{w}}_i = \CalA^T\CalG^T \widetilde{\Bs{\psi}}_i + \CalA^T\left(I_{MN} - \CalG^T\right)\Bs{w}^\circ {.}
\end{align} 
Finally, substituting \eqref{eq:errorpsi} into the above expression we arrive at the following error recursion:
\begin{align}
\blderr{w}_i = \BCalB_i \blderr{w}_{i-1} - \Bs{s}_i + \Bs{r} {,} \label{eq:errmodel}
\end{align}
where 
\begin{align}
\BCalB_i &= \CalA^T \CalG^T (I_{MN} - \CalM \BCalR_{u,i}) {,} \label{eq:matrixBi}\\
\BCalR_{u,i} &= \Diag{(\bu_{k,i}\bu_{k,i}^T)^N_{k=1}} {,} \label{eq:calRui}\\
\CalM &= \Diag{(\mu_k I_M)^N_{k=1}} {,} \label{eq:calM} \\
\Bs{s}_i &= \CalA^T \CalG^T \CalM \Col{(\bu_{k,i} \bv_{k,i})^N_{k=1}} {,}\\
\Bs{r} &= \CalA^T(I_{MN} - \CalG)^T \Bs{w}^\circ {.} \label{eq:r}
\end{align}

\subsection{Mean Error Analysis}\label{ssec:meanstability}
Suppose Assumptions \ref{asmp:regressor}-\ref{asmp:para} all hold, then by taking expectation on both sides of \eqref{eq:errmodel}, we obtain:
\begin{align}
\E \blderr{w}_i = \CalB \E\blderr{w}_{i-1} + r, \label{eq:meanrecursion}
\end{align}
where
\begin{align}
\CalB &= \E\BCalB_i = \CalA^T \CalG^T (I_{MN} - \CalM \CalR_u), \label{eq:matrixB} \\
r &=\E\Bs{r}\;\; =\CalA^T(I_{MN}-\CalG)^T\E\Bs{w}^\circ {,}\label{eq:Er}
\end{align}
and 
\begin{align}
\CalR_u = \E\BCalR_{u,i}=\Diag{(R_{u,k})^N_{k=1}}. \label{eq:calRu}
\end{align}
\begin{Theorem}
\label{thm:mean}
{\rm (Mean stability)} Suppose that Assumptions \ref{asmp:regressor}-\ref{asmp:para} hold. Then, MAIC is stable in the mean if the step-size $\mu_k$ is chosen such that 
\begin{align}
\mu_k < \frac{2}{\lambda_{\max} (R_{u,k})} \label{eq:mean_mu}.
\end{align}
Furthermore, if Assumption \ref{asmp:mean} also holds, then MAIC is asymptotically unbiased.
\end{Theorem} 
\begin{IEEEproof}
See Appendix \ref{sec:appdx_A}.
\end{IEEEproof}

From Theorem \ref{thm:mean}, we see that condition \eqref{eq:mean_mu} is also the one that ensures the mean stability of a single node in the non-cooperative case \cite{SYT12TSP,SYT13SPM,AHS14Proc}. Therefore, if individual nodes are stable in the mean, the network that applies MAIC is also stable in the mean as well. By contrast, even though every individual node is stable in the mean in the non-cooperative case, MDLMS is not guaranteed to be mean stable since its mean stability depends on the regularization weight $\eta$ \cite{JC14TSP}.

\subsection{Mean-Square Error Analysis}\label{ssec:meansquarestability}
We next study the mean-square stability of the proposed MAIC strategy, under the Assumptions \ref{asmp:regressor}-\ref{asmp:para}. From the error recursion \eqref{eq:errmodel}, we have for any compatible non-negative definite matrix $\Sigma$,
\begin{dmath}
\Norm{\blderr{w}_i}^2_\Sigma 
= \blderr{w}^T_{i-1}\BCalB^T_i\Sigma\BCalB_i\blderr{w}_{i-1} + \Bs{s}^T_i \Sigma \Bs{s}_i +  \Bs{r}^T \Sigma \Bs{r} 
+ 2\Bs{r}^T\Sigma\BCalB_i\blderr{w}_{i-1} - 2\blderr{w}^T_{i-1}\BCalB_i^T\Sigma\Bs{s}_i - 2\Bs{s}^T_i\Sigma\Bs{r} {,} \label{eq:errsqnorm}
\end{dmath}
where matrix $\BCalB_i$ is given in \eqref{eq:matrixBi}. Taking expectation on both sides of the expression \eqref{eq:errsqnorm}, the last two terms of \eqref{eq:errsqnorm} evaluate to zero according to the Assumptions \ref{asmp:regressor}-\ref{asmp:para}, therefore we have the following variance relation:
\begin{dmath}
\E\Norm{\blderr{w}_i}^2_\Sigma =
\E\Norm{\blderr{w}_{i-1}}^2_{\Sigma'} + \E\Norm{\Bs{s}_i}^2_\Sigma + \E\Norm{\Bs{r}}^2_\Sigma + 2\E (\Bs{r}^T\Sigma\BCalB_i\blderr{w}_{i-1})
{,} \label{eq:errmeansqnorm}
\end{dmath}
where 
\begin{align}
\Sigma'= \E \left[ \BCalB^T_i\Sigma\BCalB_i\right] {.} \label{eq:Sigmaprime}
\end{align}
Letting $\sigma = \Vc(\Sigma)$ and evaluating the second term of the R.H.S.\ of \eqref{eq:errmeansqnorm}, we have 
\begin{align}
\E\Norm{\Bs{s}_i}^2_\Sigma 
&= \E\Tr\left(\Bs{s}_i\Bs{s}^T_i\Sigma\right) \nonumber\\
&= \Tr\left[\left( \E\Bs{s}_i\Bs{s}^T_i\right) \Sigma\right] \nonumber\\
&=\Tr\left(\CalA^T\CalG^T\CalM\CalS\CalM\CalG\CalA \Sigma\right)  \nonumber\\
&= \Vc\left(\CalA^T\CalG^T\CalM\CalS\CalM\CalG\CalA\right)^T \sigma , \label{eq:errmeansqnorm2nd} 
\end{align} 
where 
\begin{align}
\CalS = \Diag{(\sigma^2_{v,k} R_{u,k})^N_{k=1}}, \label{eq:matrixS}
\end{align}
and the equality \eqref{eq:errmeansqnorm2nd} follows from the identity $\Tr(AB) = \Vc(A^T)^T\Vc(B)$. Using a similar argument, we also have
\begin{align}
\E\Norm{\Bs{r}}^2_\Sigma = \Vc \left[\CalA^T(I_{MN} - \CalG)^T\CalR_w(I_{MN} - \CalG)\CalA\right]^T\sigma. \label{eq:errmeansqnorm3rd}
\end{align}
As for the last term of \eqref{eq:errmeansqnorm}, we obtain
\begin{align}
\E\left[\Bs{r}^T\Sigma\BCalB_i\blderr{w}_{i-1}\right] 
&= \E\left[\Vc\left(\Bs{r}^T\Sigma\BCalB_i\blderr{w}_{i-1}\right)\right]  \nonumber\\
&= \E\left[\left(\BCalB_i\blderr{w}_{i-1}\right)^T \otimes \Bs{r}^T\right] \sigma \label{eq:errmeansqnorm4th}\\
&= \E\left[(\BCalB_i \blderr{w}_{i-1}) \otimes (I_{MN}\cdot\Bs{r}) \right]^T \sigma \nonumber\\
&= \left[\E (\BCalB_i\otimes I_{MN}) \E(\blderr{w}_{i-1}\otimes \Bs{r})\right]^T\sigma \label{eq:BK}\\
&= \left[\CalB_I \E(\blderr{w}_{i-1}\otimes \Bs{r})\right]^T\sigma, \label{eq:BK2}
\end{align}
where
\begin{align}
\CalB_I=\E (\BCalB_i\otimes I_{MN})=\CalB\otimes I_{MN}. \label{eq:BI}
\end{align}
We have used the identity $\Vc(AXB)=(B^T\otimes A)\Vc(X)$ to obtain \eqref{eq:errmeansqnorm4th}, and \eqref{eq:BK} follows from identity $(AC)\otimes(BD)=(A\otimes C)(B\otimes D)$. Let $\sigma'=\Vc(\Sigma')$, so that from \eqref{eq:Sigmaprime}, we have
\begin{align}
\sigma' = \CalE \sigma, \label{eq:Esigma}
\end{align}
where
\begin{align} 
\CalE &= \E\left[ \BCalB_i^T \otimes \BCalB_i^T \right] \nonumber \\
&= [ I_{M^2N^2} - I_{MN}\otimes\CalM\CalR_u - \CalM\CalR_u\otimes I_{MN} 
+ \left( \CalM\otimes\CalM\right)  \E\left(\BCalR_{u,i}\otimes\BCalR_{u,i}\right) ] \; \CalG\CalA\otimes\CalG\CalA . \label{eq:matrixE}
\end{align}
From \eqref{eq:errmeansqnorm}, \eqref{eq:errmeansqnorm2nd}, \eqref{eq:errmeansqnorm3rd}, \eqref{eq:BK2}, and \eqref{eq:Esigma}, we obtain
\begin{align}
\E\Norm{\blderr{w}_i}^2_\sigma =
\E\Norm{\blderr{w}_{i-1}}^2_{\CalE\sigma} + \left[f_a + f_b + f_{c,i-1}\right]^T\sigma, \label{eq:variancerelation}
\end{align}
where
\begin{align}
f_a &= \Vc \left(\CalA^T\CalG^T\CalM\CalS\CalM\CalG\CalA\right), \\
f_b &= \Vc \left[\CalA^T(I_{MN} - \CalG)^T\CalR_w(I_{MN} - \CalG)\CalA\right], \label{eq:gb}\\
f_{c,i-1} &= 2\CalB_I \E(\blderr{w}_{i-1}\otimes \Bs{r}). \label{eq:gc}
\end{align}
The following result follows almost immediately. 

\begin{Theorem}\label{thm:means_quare}
{\rm (Mean-Square Stability)}
Suppose that Assumptions \ref{asmp:regressor}-\ref{asmp:para} hold, and the matrix $\CalB$ is stable, i.e., $\rho(\CalB) < 1$. Then, MAIC is stable in the mean-square sense if and only if the matrix $\CalE$ is stable (i.e., $\rho(\CalE) < 1$). 
\end{Theorem}
\begin{IEEEproof}
See Appendix \ref{sec:appdx_B}.
\end{IEEEproof}

To obtain the steady-state network MSD using $\CalE$ directly makes the analysis intractable. Therefore, in the following, we derive an approximation for $\CalE$, and use that to derive an approximation for the steady-state network MSD. This approach follows the practice commonly adopted in various works like \cite{CAT10TSP,AHS12TR,JC14TSP} when analyzing mean-square stability of diffusion strategies. Recalling \eqref{eq:matrixB}, we let
\begin{align}
\CalF
&=\CalB^T\otimes\CalB^T \nonumber\\
&= [I_{M^2N^2} - I_{MN}\otimes\CalM\CalR_u - \CalM\CalR_u\otimes I_{MN} 
+ \left( \CalM\otimes\CalM\right)  \left(\CalR_u\otimes\CalR_u\right)] \; \CalG\CalA\otimes\CalG\CalA. \label{eq:matrixF}
\end{align}
Comparing \eqref{eq:matrixE} and \eqref{eq:matrixF}, we see that $\CalE - \CalF = \cO(\CalM^2)$ and 
\begin{align}
\E\Norm{\blderr{w}_{i-1}}^2_{\CalE\sigma} = \E\Norm{\blderr{w}_{i-1}}^2_{\CalF\sigma} + \cO(\Norm{\CalM}^2)\E\Norm{\blderr{w}_{i-1}}^2_{\sigma} . \label{eq:errmeansqnorm1stapprx}
\end{align}
When the recursion \eqref{eq:variancerelation} is stable, i.e., MAIC is mean-square stable, the last term in \eqref{eq:errmeansqnorm1stapprx} is negligible when the step-sizes $\{\mu_k\}_{k=1}^N$ are chosen to be sufficiently small since $\E\Norm{\blderr{w}_{i-1}}^2$ is bounded. Therefore, by adopting the approximation $\CalE\approx\CalF$, we have $\E\Norm{\blderr{w}_{i-1}}^2_{\CalE\sigma} \approx \E\Norm{\blderr{w}_{i-1}}^2_{\CalF\sigma}$, and the following approximate recursion relationship
\begin{align}
\E\Norm{\blderr{w}_i}^2_\sigma =
\E\Norm{\blderr{w}_{i-1}}^2_{\CalF\sigma} + \left[f_a + f_b + f_{c,i-1}\right]^T\sigma. \label{eq:approxvariancerelation}
\end{align}
The above recursion \eqref{eq:approxvariancerelation} is mean-square stable if and only if $\CalF$ is stable, which is achieved if and only if $\CalB$ is stable as $\rho(\CalF)=\rho(\CalB)^2$. We have from \eqref{eq:approxvariancerelation}, 
\begin{align}
\lim_{i\to\infty} \E\Norm{\blderr{w}_i}^2_{(I_{M^2N^2}-\CalF)\sigma} =\left[f_a+f_b+f_{c,\infty}\right]^T\sigma {,}
\end{align}
where
\begin{align}
f_{c,\infty} = 2\left[\CalB(I_{MN}-\CalB)^{-1}\otimes I_{MN}\right]f_b {,} \label{eq:gcstadystate}
\end{align}
which is obtained by substituting \eqref{eq:inf_kron_err} in Appendix~\ref{sec:appdx_C} into \eqref{eq:gc}.
By choosing $\sigma = (I_{M^2N^2}-\CalF)^{-1}\Vc\left(I_{MN}\right)/N$, the steady-state network MSD,
\begin{align}
\zeta = \inflim{i}\frac{1}{N}\E \Norm{\blderr{w}_i}^2,
\end{align} 
can be approximated as
\begin{align}
\zeta \approx \frac{1}{N}\left[f_a+f_b+f_{c,\infty}\right]^T\left(I_{M^2N^2} - \CalF \right)^{-1}\Vc(I_{MN}) 
{.} \label{eq:MSD} 
\end{align}

\section{Optimizing The Inter-cluster Cooperation}\label{sec:Weights}
In this section, we formulate an optimization problem to obtain the inter-cluster cooperation weights that minimize an upper bound of \eqref{eq:MSD}. For simplicity, we assume uniform step-sizes, i.e., $\mu_k = \mu$ for all $k$, throughout this section. We show that this problem is decomposable into local optimization problems at each node, and we provide an adaptive implementation of our approach.

The averaging rule \eqref{eq:averagerule} associates identical weights for inter-cluster cooperation regardless of the noise and data profiles across the nodes as well as the correlation between cluster parameters. Although the MSD performance of clusters with highly correlated parameters may be improved via cooperation with large weights, this may not be beneficial for clusters with less correlated parameters, and may even lead to performance deterioration (an illustrative example is provided in Section \ref{ssec:simB}), which in turn worsens the network MSD performance. Therefore, our aim is to optimize the inter-cluster cooperation weights in \eqref{eq:gamma_condition}, so that the cooperation between nodes from different clusters can be tuned to improve the network MSD performance. 

\subsection{Inter-cluster Weights Optimization}\label{ssec:distributed}
The basic idea of our inter-cluster weights selection scheme is to choose the weights via optimization of \eqref{eq:MSD}, which takes into account both the noise and data profile of each node and the correlation between different parameters. Since directly optimizing \eqref{eq:MSD} is computationally difficult, we instead find the inter-cluster cooperation weights to minimize a proxy of \eqref{eq:MSD}. Specifically, we drop the $f_{c,\infty}$ term in \eqref{eq:MSD}, and consider
\begin{align}
\widehat{\zeta} = \frac{1}{N}\left(f_a+f_b\right)^T\left(I_{M^2N^2} - \CalF \right)^{-1}\Vc(I_{MN}) \label{eq:MSDapprox} 
\end{align}
instead. Numerical studies in Section~\ref{sec:Sim} suggest that this approximation does not deviate from the actual MSD \eqref{eq:MSD} significantly.

It can be shown that when a uniform step-size is applied, i.e., $\mu_k = \mu$ for all $k$, an upper bound for the $\widehat{\zeta}$ in \eqref{eq:MSDapprox} is given by
\begin{dmath}
c\cdot\Tr \bigg[\mu^2\CalA^T\CalG^T\CalS\CalG\CalA + \CalA^T(I_{MN} - \CalG)^T\CalR_w(I_{MN} - \CalG)\CalA\bigg] {.} \label{eq:upperbound_cent}
\end{dmath}
where $c$ is a positive constant (see Appendix~\ref{sec:appdx_bound} for the derivation). By focusing on the terms that are dependent on the inter-cluster cooperation weights matrix $G$, \eqref{eq:upperbound_cent} can be re-written as
\begin{align}
c\cdot\left[ \Vc(\CalG)^T \mc{K} \Vc(G) - \kappa^T \Vc(\CalG)\right] ,
\end{align}
where
\begin{align}
\mc{K} &= \left[ \CalA \CalA^T \otimes (\mu^2\CalS+\CalR_w)\right], \nonumber\\
\kappa &= 2 \Vc\left[\CalR_w \CalA \CalA^T \right] \nonumber {.}
\end{align}
Let $y=\Vc(\CalG)$, 
the optimal inter-cluster cooperation weights can be obtained by solving the following centralized optimization problem, which is a quadratic programming problem \cite{BOYD2004convex}:
\begin{align*}
\text{(P1)} \; \min_{y} \quad & y^T\mc{K}y - \kappa^Ty\\
\text{s.t.} \quad &g_{\ell k} \ge 0, \; {\sum \limits_{\ell\in\NI^+}} \; {g_{\ell k} = 1},\;{g_{\ell k} = 0}, \; \text{if} \; \ell \notin \NI^+. 
\end{align*}

Although the centralized method can provide better optimization results in general, having a centralized controller to compute the inter-cluster cooperation weights is not practical for some networks. Therefore, in the sequel, we focus on a distributed optimization procedure based on a different upper bound of \eqref{eq:MSD} (see Appendix~\ref{sec:appdx_bound}):
\begin{align}
\bar{c} \cdot \Tr \left[\mu^2\CalG^T\CalS\CalG + (I_{MN} - \CalG)^T\CalR_w(I_{MN} - \CalG)\right], \label{eq:upperbound_dist}
\end{align}
where $\bar{c}$ is a positive constant. To obtain the bound \eqref{eq:upperbound_dist}, we assume that the intra-cluster combination matrix $A$ is symmetric and doubly-stochastic\footnote{A doubly stochastic matrix $X$ satisfies $X^T\mathds{1}_N=X\mathds{1}_N=\mathds{1}_N$.}. This assumption seems to be restrictive, however combination rules like the \emph{Metropolis rule} (see Section \ref{sec:Sim}) \cite{AHS12TR} which leads to symmetric and doubly-stochastic combination matrices, are computationally convenient and lead to fairly good performance compared with non-cooperative strategies \cite{SYT13SPM}. Ignoring all terms in \eqref{eq:upperbound_dist} independent of the inter-cluster weights $\{g_\lk\}$, we obtain
\begin{dmath}\label{eq:optbound}
\mu^2\sum_{k=1}^{N} \bigg\{ \sum_{\ell=1}^{N} \Big[ \mu^2 g_{\lk}^2\sigma^2_{v,\ell}\Tr(R_{u,\ell}) - 2 g_\lk \Tr(R_{w,\lk}) + g_{\ell k} \sum_{m=1}^{N} g_{mk}\Tr (R_{w,\ell m}) \Big] \bigg\} ,
\end{dmath}
where
\begin{align}
R_{w,\ell k} = \E\Bs{w}^\circ_\ell {\Bs{w}^\circ_k}^T {,} \label{eq:Rwlk}
\end{align} 
which is the $(\ell,k)$-th block of matrix $\CalR_w$. To optimize \eqref{eq:optbound}, we can decompose it into $N$ separate local optimization problems for each node $k$ as follows: 
\begin{align*}
\text{(P2)} \; \min_{q_k} \quad & q_k^T (\Omega_{a,k} + \Omega_{b,k}) q_k - 2 h_k^T q_k\\
\text{s.t.} \quad& \mathds{1}_{n_k}^T q_k = 1, \quad q_k \ge 0_{n_k}.
\end{align*}
\begin{align}
q_k &= \Col{(g_{\ell k})_{\ell\in\NI^+}}, \label{eq:qk} \\
h_k &= \Col{(\Tr(R_{w,\ell k}))_{\ell\in\NI^+}}, \\
\Omega_{a,k} &= \Diag{\big(\mu^2\sigma^2_{v,\ell}\Tr(R_{u,\ell})\big)_{\ell\in\NI^+}}, \label{eq:Omega_ak}\\
\Omega_{b,k} &= \E\big[\Bs{W}_{k}^T \Bs{W}_{k}\big], \label{eq:Omega_bk}\\
\Bs{W}_k &= [\Bs{w}^\circ_\ell]_{\ell\in\NI^+}, \label{eq:Wk}\\
n_k &= \big|\NI^+ \big| \label{eq:nk},
\end{align}
where $[\Bs{w}^\circ_\ell]_{\ell\in\NI^+}$ denotes a matrix whose columns are $\Bs{w}^\circ_\ell$ for $\ell\in\NI^+$. The problem (P2) is a quadratic programming problem, which can be solved independently using standard techniques by each node $k$ \cite{BOYD2004convex}.

\subsection{Adaptive Implementation of the Distributed Method}\label{ssec:adpt_imp}
In Section \ref{ssec:distributed}, we have implicitly assumed that the data profiles $\sigma^2_{v,k}\Tr(R_{u,k})$, for $k=1,\ldots,N$, are known, which may not be the case in practical applications. We now present an online estimation procedure similar to those in \cite{AHS12TR,XCZ12TSP,CKYU13ICASSP,JFERNANDEZ14ICASSP} to \emph{adaptively} estimate these statistics and update the inter-cluster cooperation weights. For each $\ell\in\NC$, let $\widehat{\bx}_{\ell k,i}$ be the estimate of $\mu^2\sigma^2_{v,\ell}\Tr(R_{u,\ell})$ made by node $k$ at time instant $i$. Then $\widehat{\bx}_{\ell k,i}$ can be updated recursively by the following moving-average method:
\begin{align}
\widehat{\bx}_{\ell k,i} = \alpha\cdot\widehat{\bx}_{\ell k,i-1} + (1-\alpha)\cdot\Norm{\bpsi_{\ell,i}-\bw_{k,i-1}}^2, \label{eq:xhat}
\end{align}
where the coefficient $\alpha$ is chosen from $[0,1]$. 

Although in some applications the statistics $\CalR_w$ can be obtained by probing the underlying environment before the network is deployed, we are interested in the cases that the $\CalR_w$ is not accessible and therefore both $R_{w,\lk}$ and $\Tr\left( R_{w,\ell k}\right)$ are unknown for all $\ell$ and $k$. This is more practical because the correlation information across the parameters may not be time-invariant, hence the network should be able to adapt to changes in $\CalR_w$. For realizing the adaptability over $\CalR_w$, at time instant $i$, each node uses 
\begin{align}
\widehat{\btau}_{\ell k,i} = \bw_{\ell, i-1}^T \bw_{k,i-1} \label{eq:tauhat}
\end{align}
as an instantaneous approximation for $\Tr\left(R_{w,\ell k} \right)$. Finally, the matrix $\Omega_{b,k}$ can be approximated at each time instant $i$ by 
\begin{align}
\widehat{\bOmega}_{b,k,i}=\widehat{\bW}_{k,i}^T \widehat{\bW}_{k,i} , \label{eq:hatOmega}
\end{align}
where $\widehat{\bW}_{k,i}=[\bw_{\ell, i-1}]_{\ell\in\NI^+}$. Having $\widehat{\bx}_{\lk,i}$, $\widehat{\btau}_{\lk,i}$, and $\widehat{\bOmega}_{b,k,i}$ obtained by \eqref{eq:xhat}, \eqref{eq:tauhat}, and \eqref{eq:hatOmega} respectively, the distributed weights optimization method proposed in Section \ref{ssec:distributed} can then be implemented adaptively as summarized in Algorithm~\ref{alg:alg1}.

\begin{algorithm}[!ht]
	\caption{MAIC with adaptive update of inter-cluster cooperation weights}\label{alg:alg1}
	\begin{algorithmic}[1] 
		\STATE {Initialized with\\ $w_{k,0}=0_M$, $\widehat{\bx}_{\lk,0}=0$ for all $k$, weights $\hat{g}_{\lk,0}=0$.}
		\FOR {every node $k$ at each time instant $i\ge1$}
		\STATE {$\bpsi_{k,i} = \bw_{k,i-1} + \mu_k \bu_{k,i} ( \bd_{k,i} - \bu_{k,i}^T \bw_{k,i-1})$}
		\STATE {Receive $\bpsi_{\ell,i}$ from all $\ell\in\NC$ and compute:\\
			$\widehat{\bx}_{\lk,i}=\alpha\cdot\widehat{\bx}_{\lk,i-1} + (1-\alpha)\cdot\Norm{\bpsi_{\ell,i}-\bw_{k,i-1}}^2$
		\STATE {Receive $w_{\ell,i-1}$ from all $\ell\in\NI^+$ and update:\\
			$\widehat{\btau}_{\lk,i} = \bw_{\ell, i-1}^T \bw_{k,i-1}$} \\
			$\widehat{\bW}_{k,i}=[\bw_{\ell, i-1}]_{\ell\in\NI^+}$ }
		\STATE {Update the following quantities:\\ 
			$\widehat{\bh}_{k,i}=\Col{(\widehat{\btau}_{\lk,i})_{\ell\in\NI^+}}$ \\
			$\widehat{\bOmega}_{a,k,i}=\Diag{(\widehat{\bx}_{\lk,i})_{\ell\in\NI^+}}$\\			
			$\widehat{\bOmega}_{b,k,i}=\widehat{\bW}_{k,i}^T \widehat{\bW}_{k,i}$\\
			$\widehat{\bOmega}_{k,i} = \widehat{\bOmega}_{a,k,i} + \widehat{\bOmega}_{b,k,i}$}
		\STATE {Solve (P2) with quantities\\
		 ${h}_k$, ${\Omega}_{a,k}$, ${\Omega}_{b,k}$, and ${\Omega}_{k}$ replaced by\\
		 $\widehat{\bh}_{k,i}$, $\widehat{\bOmega}_{a,k,i}$, $\widehat{\bOmega}_{b,k,i}$, and $\widehat{\bOmega}_{k,i}$, respectively\\ to obtain the solution $\hat{q}_{k,i} = \Col{(\hat{g}_{\lk,i})_{\ell\in\NI^+}}$}
		\STATE {Perform inter-cluster cooperation:\\
			$\bphi_{k,i} = \sum \limits_{\ell \in \NI^+} \hat{g}_{\lk,i} \bpsi_{\ell,i}$}
		\STATE {Perform intra-cluster combination:\\
			$\bw_{k,i} = \sum \limits_{\ell \in \NC} a_{\ell k}\bphi_{\ell,i}$}
		\ENDFOR
	\end{algorithmic}
\end{algorithm}

\begin{figure}[!t]
	\centering
	\subfloat[Data profile]{\includegraphics[width=2.8in]{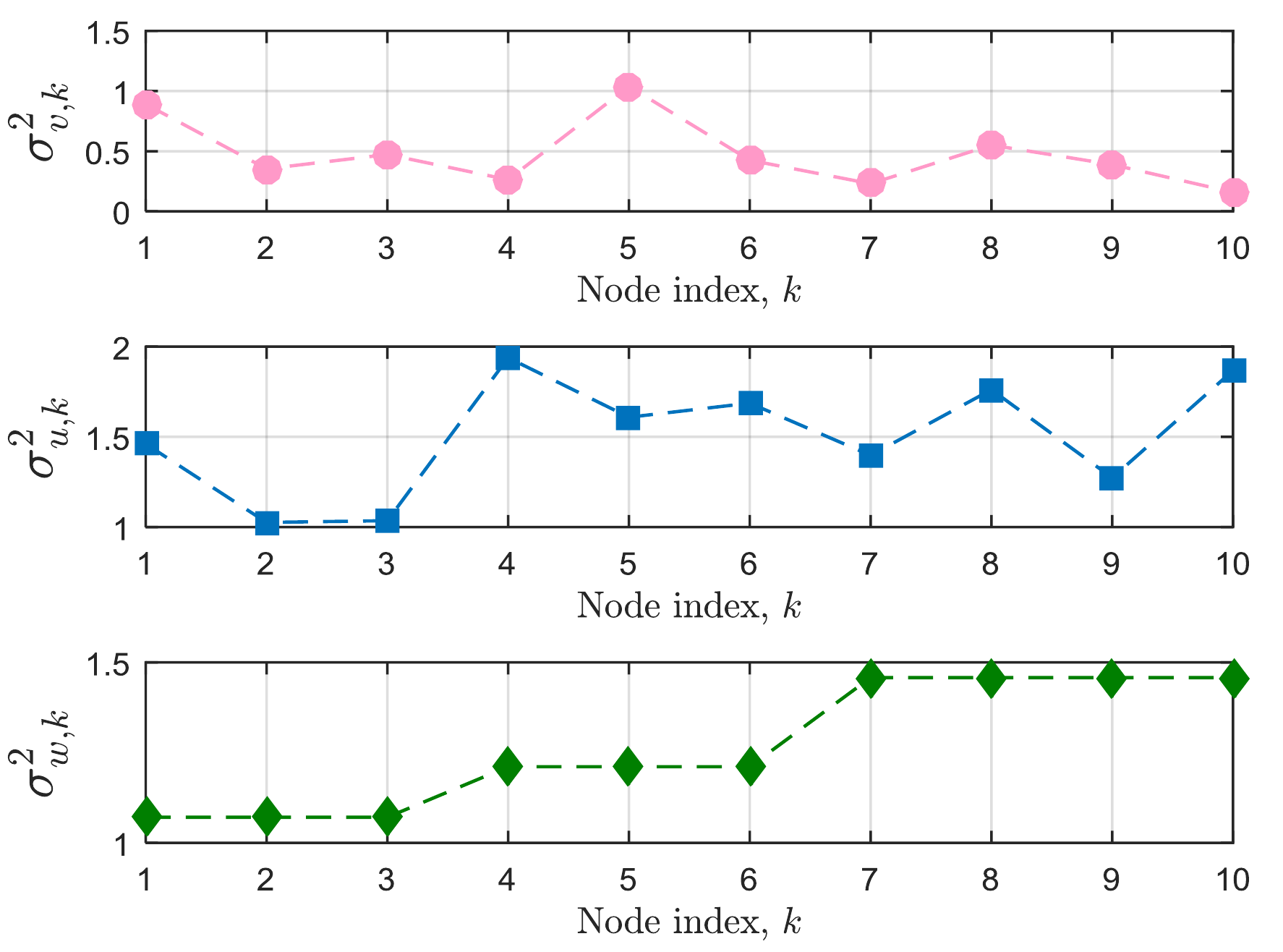}%
		\label{fig:profile}}
	\hfil
	\subfloat[Network topology]{\includegraphics[width=2.2in]{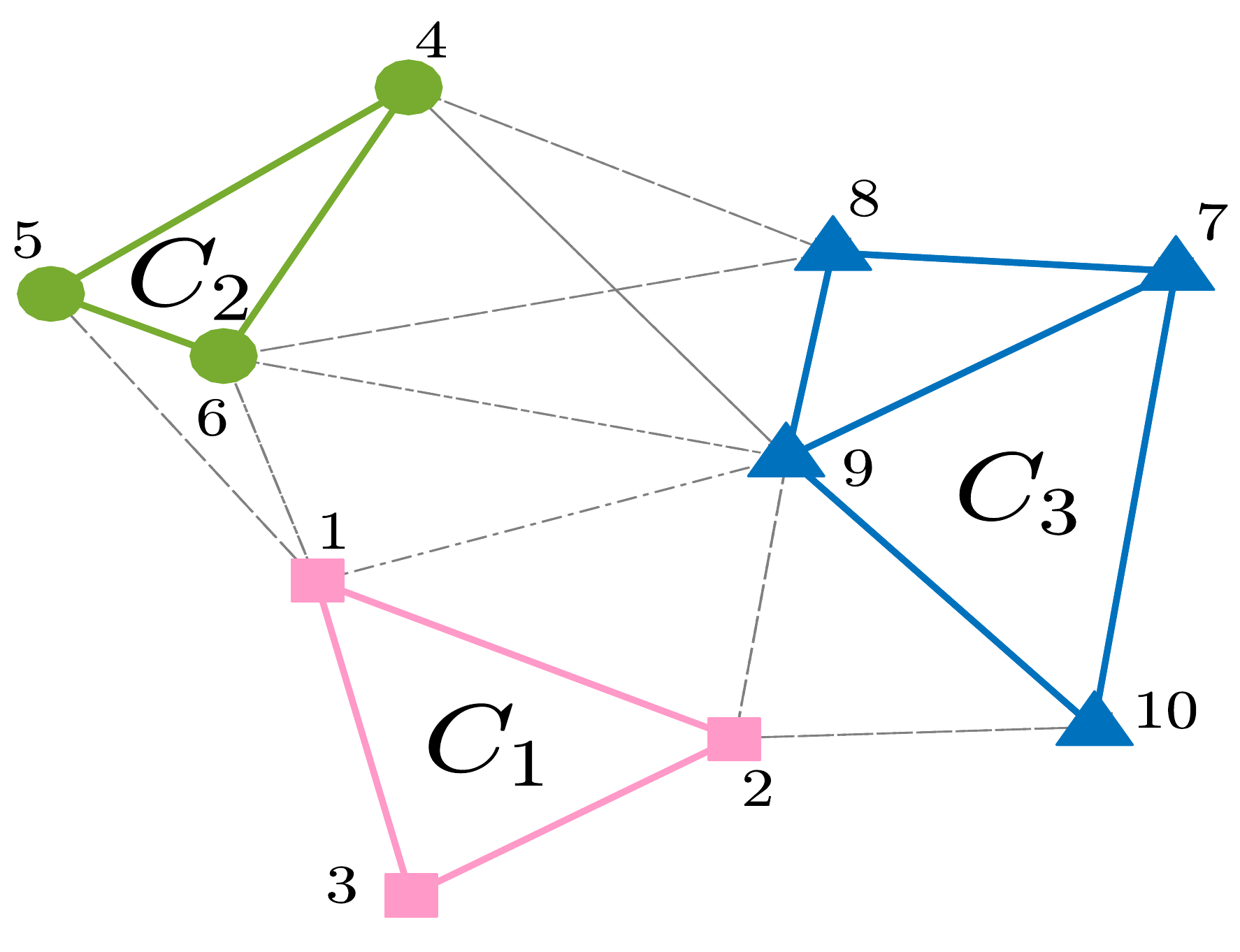}
		\label{fig:topology}}
	\hfil
	\caption{(a) Data and noise profile of each node. From top to bottom: i) noise power ii) input signal power iii) basic parameter variance. (b) The network topology. The numbers next to the nodes are node indices.}
	\label{fig:networkmodel}
\end{figure}

\section{Simulation Results}\label{sec:Sim}
In this section, we provide examples to compare the network MSD performance of MAIC to those of some other strategies in the literature. The network we tested consists of 10 nodes as depicted in Fig.~\ref{fig:topology}. The clusters $C_1$, $C_2$, and $C_3$ are denoted by pink squares, green circles, and blue triangles, respectively. The \emph{Metropolis rule} is adopted because it results in a symmetric doubly-stochastic intra-cluster combination matrix, and has low complexity and good performance in terms of both convergence rate and steady-state MSD for single-task diffusion networks \cite{AHS12TR}. The \emph{Metropolis rule} has intra-cluster combination weights given by
\begin{align}
 a_{\lk}  =
\begin{cases}\label{eq:metropolis}
\ofrac{\max\{|\NC|,|\NCell|\}},\ &\text{if }k\neq\ell\text{ and }\ell\in\NC,  \\
1-\sum\limits_{\ell \in \NC\backslash\{k\}} a_{\ell k},\ &\text{if } k=\ell,  \\ 
0,\ &\text{otherwise.}
\end{cases} 
\end{align}
All the experiment results are averaged over $10,000$ Monte-Carlo runs.

The performance of the following strategies will be compared in three different experiments in the sequel:
\begin{enumerate} [(i)]
	\item MAIC with the inter-cluster cooperation weights optimized by the centralized method in Section \ref{ssec:distributed} by solving (P1).
	\item MAIC with the inter-cluster cooperation weights optimized by the distributed method in Section \ref{ssec:distributed} by solving (P2).
	\item  MAIC with inter-cluster cooperation weights selected adaptively by Algorithm \ref{alg:alg1}.
	\item MDLMS proposed by \cite{JC14TSP} with the cooperation weights selected by the averaging rule \eqref{eq:averagerule}.
	\item MDLMS with adaptive regularization proposed in \cite{SMONAJEMI15MLSP}.
	\item The conventional ATC strategy without inter-cluster cooperation. In this case, each cluster acts as an independent subnetwork that performs ATC using the Metropolis combination rule. In \cite{JC15TSP}, a clustering strategy using adaptive adjustment of the inter-cluster cooperation weights was proposed, which leads to no cooperation between clusters with different parameters if the cluster parameters are known a priori and sensor measurements are noiseless. This comparison benchmark is then equivalent to the ideal case in \cite{JC15TSP}, and is used to avoid including errors introduced by the clustering strategy.
\end{enumerate}

\begin{figure*}[!t]
	\centering
	\subfloat[Network MSD performance comparison]{\includegraphics[width=3.0in]{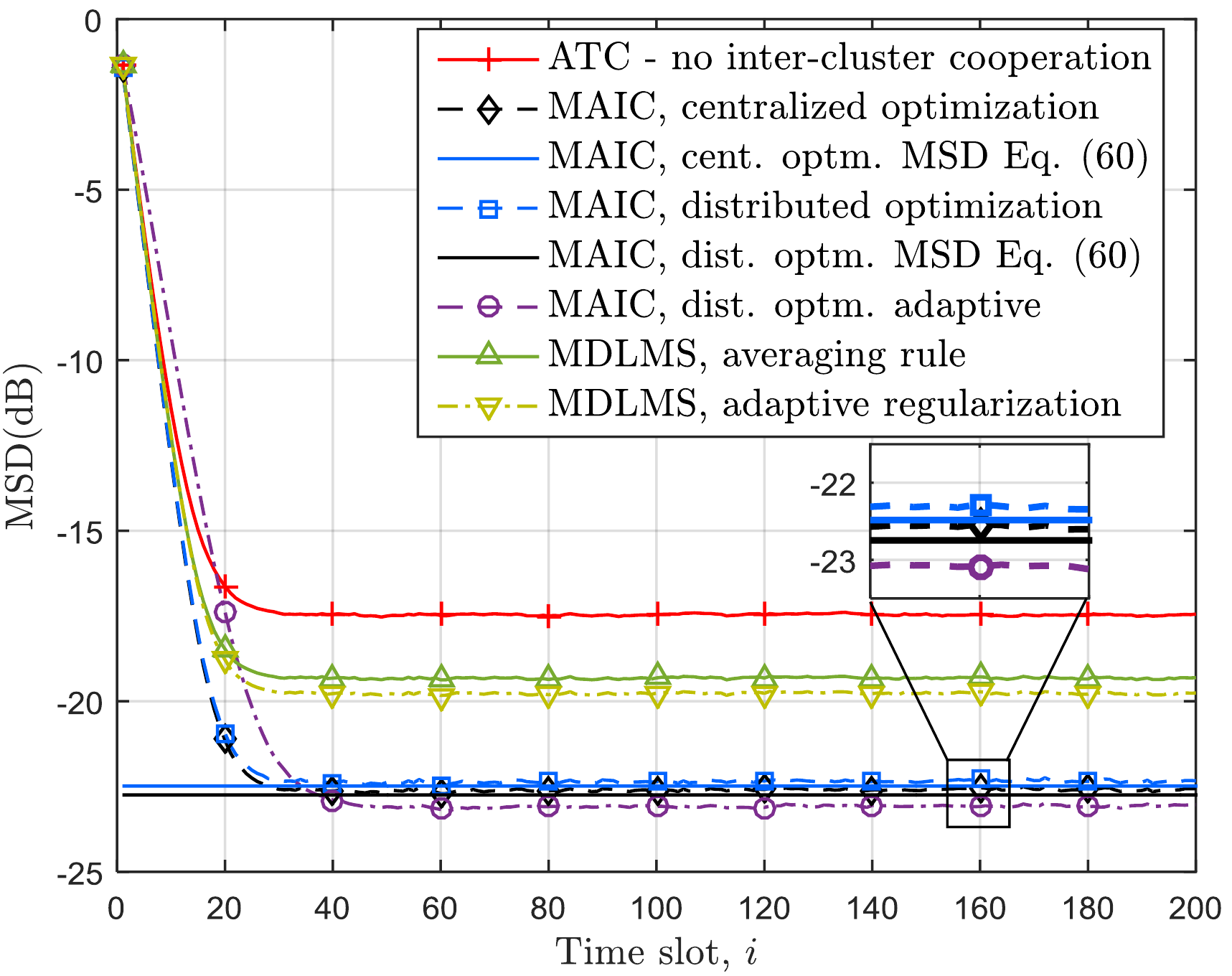}
		\label{fig:firstEg}}
	\hfil
	\subfloat[MSD model validation]{\includegraphics[width=3.0in]{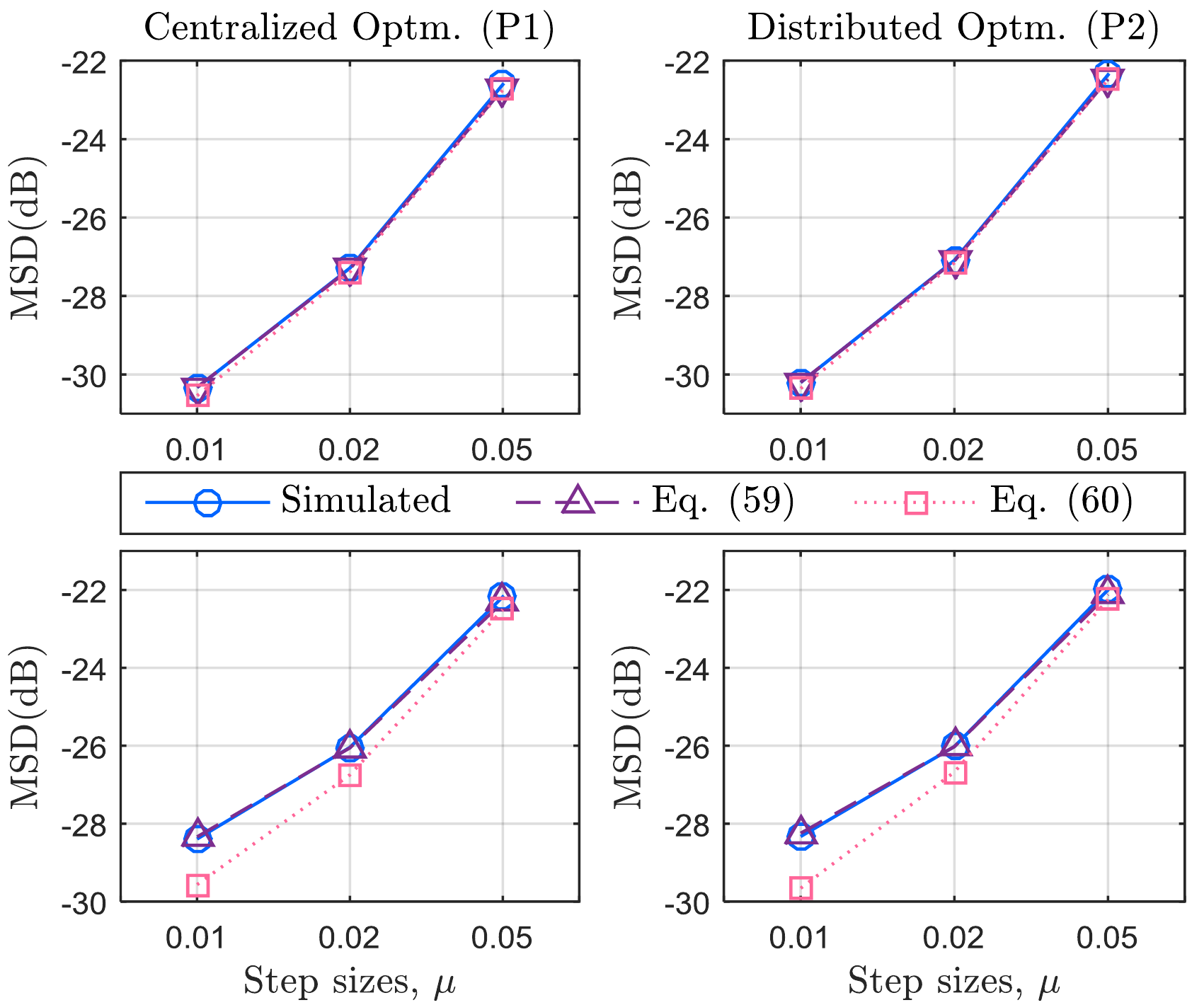}
		\label{fig:validation}}
	\hfil
	\caption{(a) Network MSD performance comparison amongst different strategies (i)-(vi). (b) Plots of simulated MSDs, theoretical MSDs (Eq.~\eqref{eq:MSD}), and approximate theoretical MSDs (Eq.~\eqref{eq:MSDapprox}) denoted by blue circles, purple triangles, and pink squares, respectively, under different experiment settings: $s_v=0.01^2$ in the upper subplots and $s_v=0.03^2$ in the lower subplots.}
	\label{fig:illustration}
\end{figure*}
\begin{figure*}
	\centering
	\subfloat[$s_v=0.01^2$]{\includegraphics[width=3.0in]{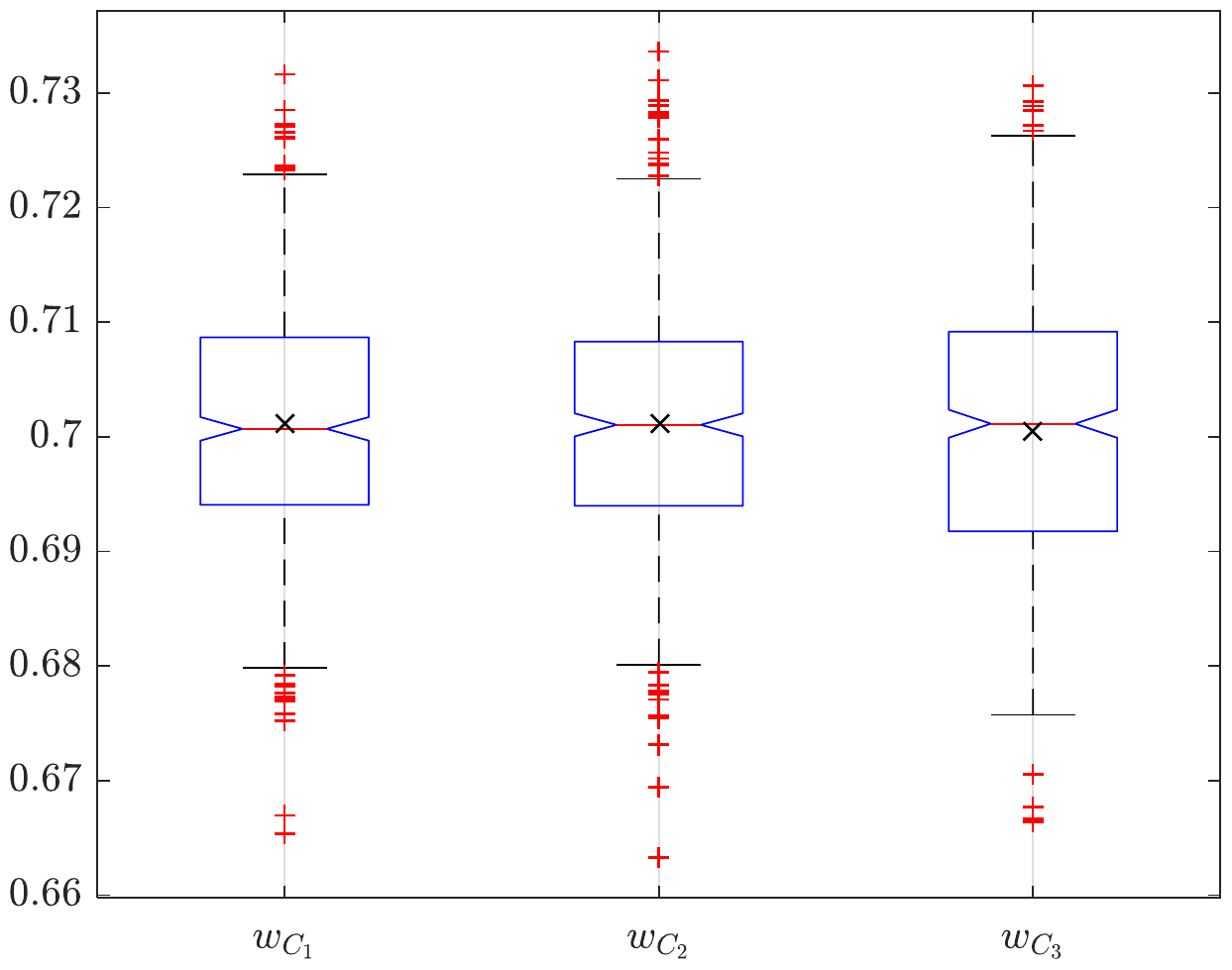}%
		\label{fig:para_box_001}}
	\hfil
	\subfloat[$s_v=0.03^2$]{\includegraphics[width=3.0in]{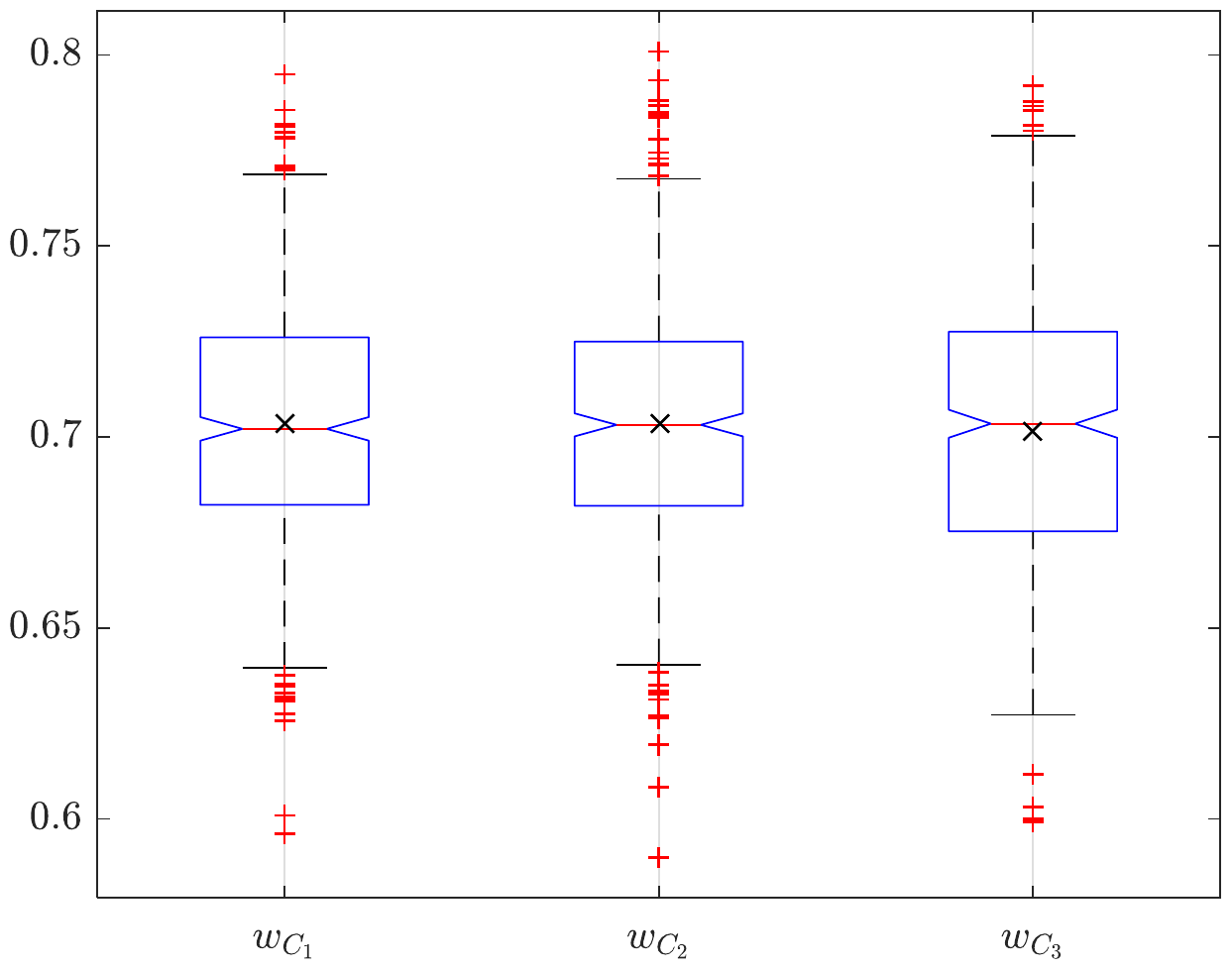}
		\label{fig:para_box_003}}
	\hfil
	\caption{Distributions of the cluster parameters $\{\Bs{w}_{C_p}^\circ\}^3_{p=1}$ under two different settings of the coefficient $s_v$: the center red bars and black crosses are the medians and means of realizations of each parameter respectively, the boxes' upper and lower edges envelope 50 percent of realizations of each parameter, red crosses outside the upper and lower black bars represent outliers. }
	\label{fig:para_box}
\end{figure*}

\subsection{Illustrative Examples}\label{ssec:simA}
The cluster parameters are chosen to be $M\times 1$ random parameters with $M=2$. The mean of each random cluster parameter is set to be $\overline{w}=0.7\times\mathds{1}_M$, and the correlation matrix of the cluster parameters is given by
\begin{align}
\begin{bmatrix}
  I_M&{0.9I_M}&{0.5I_M} \\ 
  {0.9I_M}&I_M&{0.5I_M} \\ 
  {0.5I_M}&{0.5I_M}&I_M 
\end{bmatrix}. \label{eq:corrSim}
\end{align}
The zero-mean regressor $\bu_{k,i}$ has covariance $R_{u,k} = \sigma^2_{u,k} I_M$. Fig.~\ref{fig:networkmodel}(a) shows $\sigma^2_{u,k}$ and the noise variances $\sigma^2_{v,k}$. The covariance matrix of each parameter $\Bs{w}^\circ_k$ is $s_v \sigma^2_{w,k}I_M$. We vary $s_v$ for different experiments while keeping $\sigma^2_{w,k}$ fixed as shown in Fig.~\ref{fig:networkmodel}(a). We can compute the $(\ell,k)$-th block of matrix $\CalR_w$, i.e., $\CalR_{w,\ell k} = s_v\gamma_{\ell k} \sigma_{w,\ell}\sigma_{w,k}I_M + \overline{w}\overline{w}^T$, where $\gamma_{\ell k}$ is the correlation coefficient between $\Bs{w}^\circ_\ell$ and $\Bs{w}^\circ_k$, shown as the scalar multiplier in the $(\ell,k)$ block entry in \eqref{eq:corrSim}.

In this simulation, we set $s_v=0.01^2$, and the realizations of the cluster parameters $\{\Bs{w}_{C_p}^\circ\}^3_{p=1}$ are shown in Fig.~\ref{fig:para_box}(a). The step-size is set to $\mu=0.05$ for all the nodes. The learning coefficient is set to $\alpha=0.7$ for Algorithm \ref{alg:alg1}. To make a fair comparison, we make the assumption that the cluster information is known a priori for the adaptive regularization method in \cite{SMONAJEMI15MLSP} such that node clustering is not needed and the regularization is now imposed on the inter-cluster information exchange only. In Fig.~\ref{fig:illustration}(a) and (b), we can see that the approximate theoretical steady-state network MSDs computed using \eqref{eq:MSDapprox} with inter-cluster cooperation weights obtained by the centralized optimization (P1) and the distributed optimization (P2) match well with the simulated MSDs, and do not differ significantly from the theoretical steady-state MSDs obtained from \eqref{eq:MSD}. 

Fig.~\ref{fig:illustration}(a) also shows that the MAIC strategies with optimized inter-cluster cooperation weights (i.e., strategy (i), (ii), and (iii)) achieve lower steady-state MSDs than MDLMS with cooperation weights selected by the averaging rule given in \eqref{eq:averagerule}, and MDLMS with adaptive regularization. The regularization weight used is chosen to be $\eta=1$, which minimizes the steady-state network MSD of MDLMS.

\begin{figure*}[!ht]
	\centering
	\includegraphics[width=6.1in]{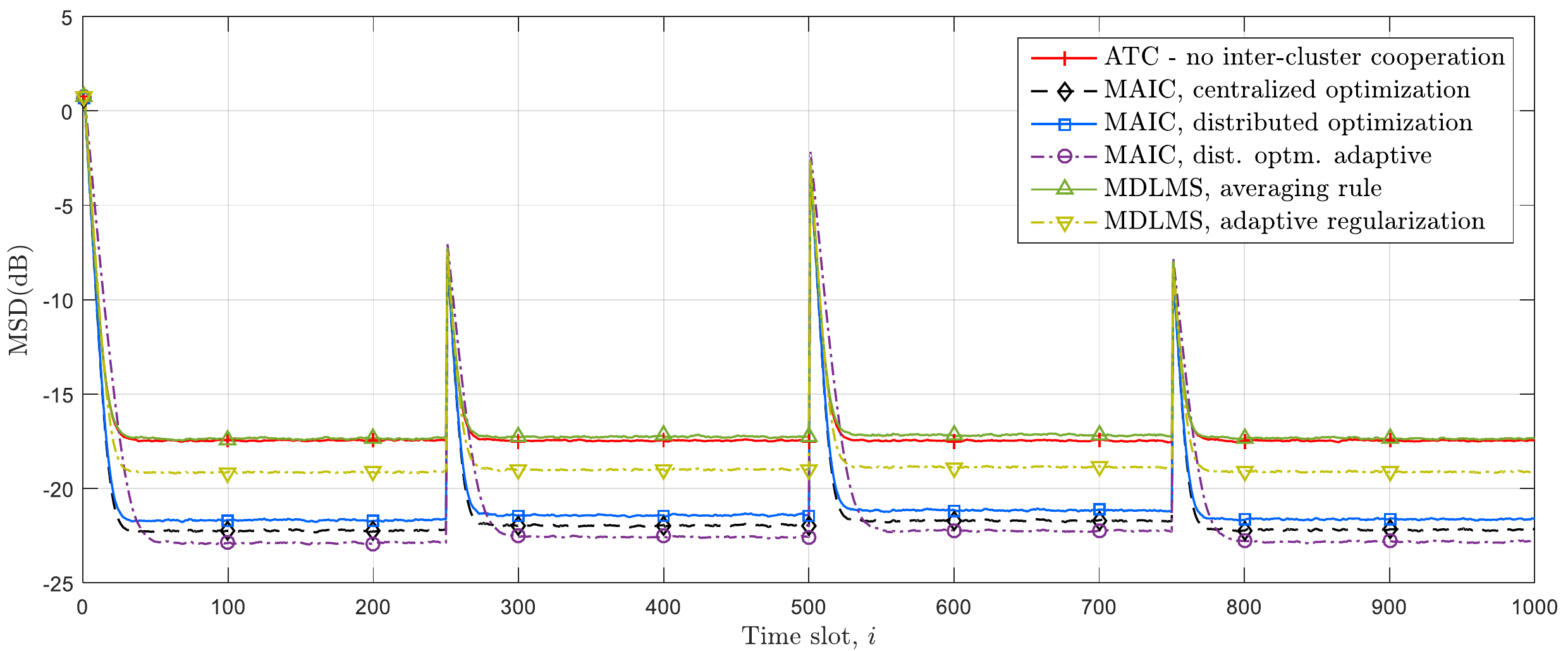}
	\caption{Network MSD comparison in the non-stationary case.}
	\label{fig:nonstationary}
\end{figure*}

In addition, we also compare the tracking performance of each strategy in a non-stationary environment where the means and correlations of the cluster parameters change at the time instants $i=250m$, for $m=1,2,3$. Specifically, we set $\overline{w}=0.8$ and \eqref{eq:corrSim} as the correlation matrix for $1\le i<250$ and $750\le i\le 1000$, and
\begin{align}
\begin{cases}
\overline{w}=0.6, \gamma_{12}=\gamma_{23}=0.5, \gamma_{13}=0.1, \text{ for } 250\le i<500,\\
\overline{w}=1.2, \gamma_{12}=\gamma_{23}=\gamma_{13}=0.1, \text{ for } 500\le i<750. \nonumber
\end{cases}
\end{align}
The regularization weight is now set to $\eta=12$ for both MDLMS with averaging rule and MDLMS with adaptive regularization. From Fig.~\ref{fig:nonstationary} we see that although MDLMS with adaptive regularization improves over its counterpart with stationary weights chosen by the averaging rule, the proposed MAIC strategies with optimized weights still achieve lower steady-state network MSDs. This is because the adaptive regularization method still needs to tune the weight $\eta$ that governs the total degree of inter-cluster cooperation as in the conventional MDLMS \cite{JC14TSP}. Therefore the improvement by MDLMS with adaptive regularization over MDLMS with averaging rule is limited. 

\subsection{Benefits of Optimized Inter-cluster Cooperation}\label{ssec:simB}

In this simulation, we set $M=1$ and the means of the cluster parameters $\{\Bs{w}_{C_p}^\circ\}^3_{p=1}$ are $\overline{w}=1$. The correlation matrix of the cluster parameters is
\begin{align}
\left[ {\begin{array}{*{20}{c}}
  1&{\gamma_{12}}&{0.5} \\ 
  {\gamma_{21}}&1&{0.5} \\ 
  {0.5}&{0.5}&1 
\end{array}} \right], \nonumber
\end{align}
\begin{figure*}[!t]
	\centering
	\subfloat[Performance of cluster $C_2$]{\includegraphics[width=3.0in]{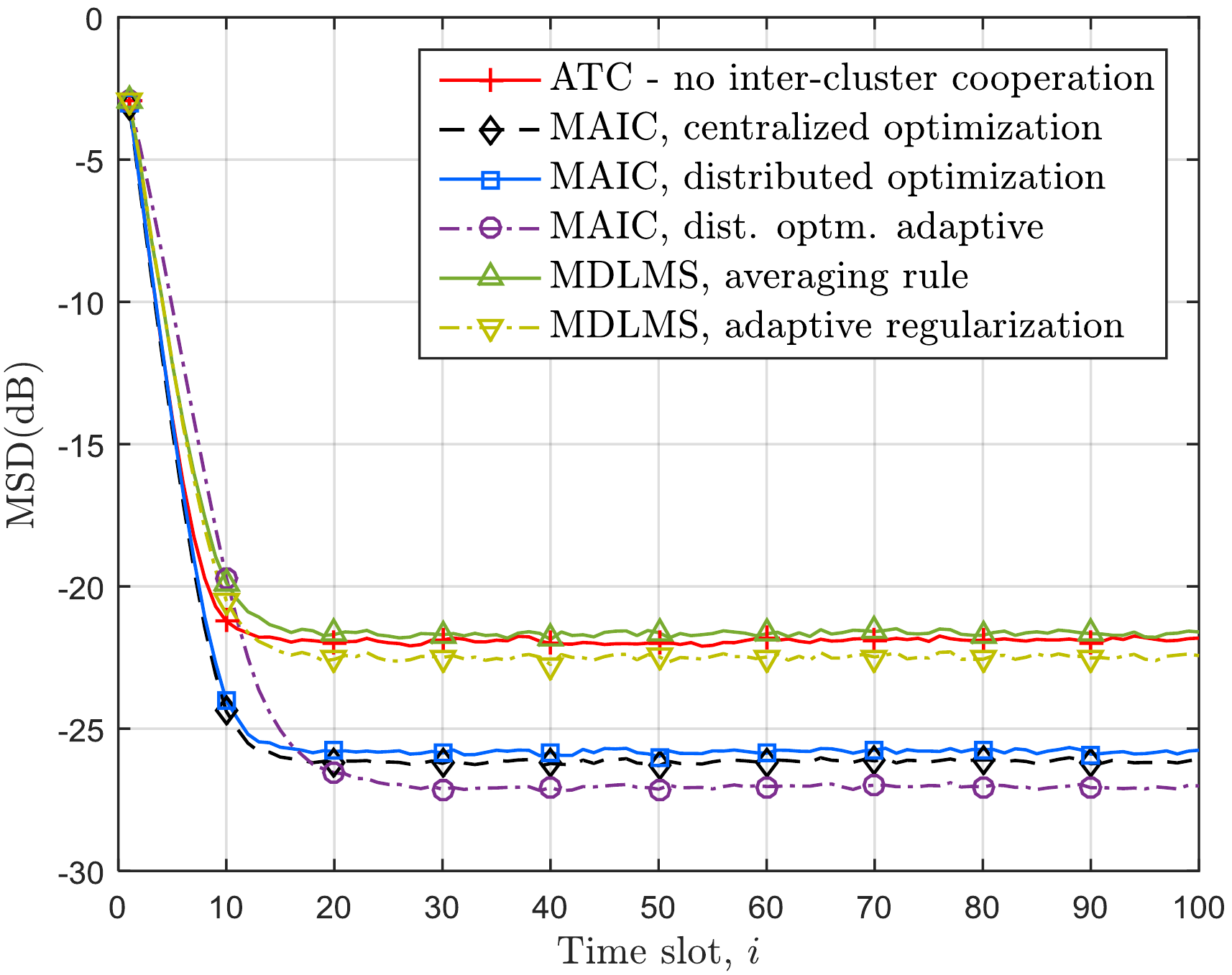}
		\label{fig:performance_C2}}
	\hfil
	\subfloat[Performance of cluster $C_3$]{\includegraphics[width=3.0in]{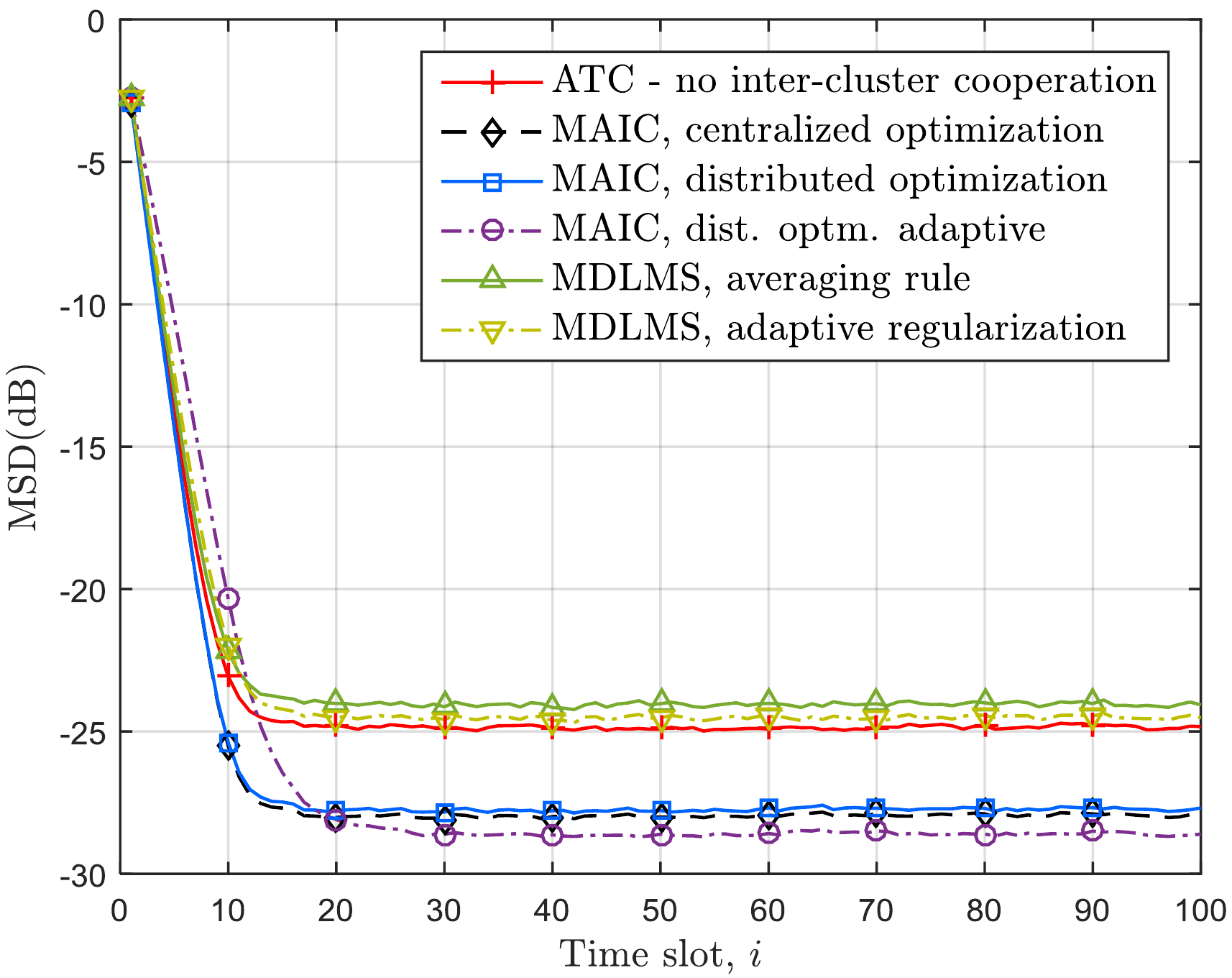}
		\label{fig:performance_C3}}
	\hfil
	\caption{Network MSD of clusters (a) $C_2$ and (b) $C_3$, when the regularization weight of MDLMS is $\eta=5$. The performance of these clusters do not benefit from the inter-cluster cooperation in MDLMS when compared with ATC with no inter-cluster cooperation.}
	\label{fig:cluster}
\end{figure*}
\begin{figure*}[!t]
	\centering
	\subfloat[Network proformance]{\includegraphics[width=3.0in]{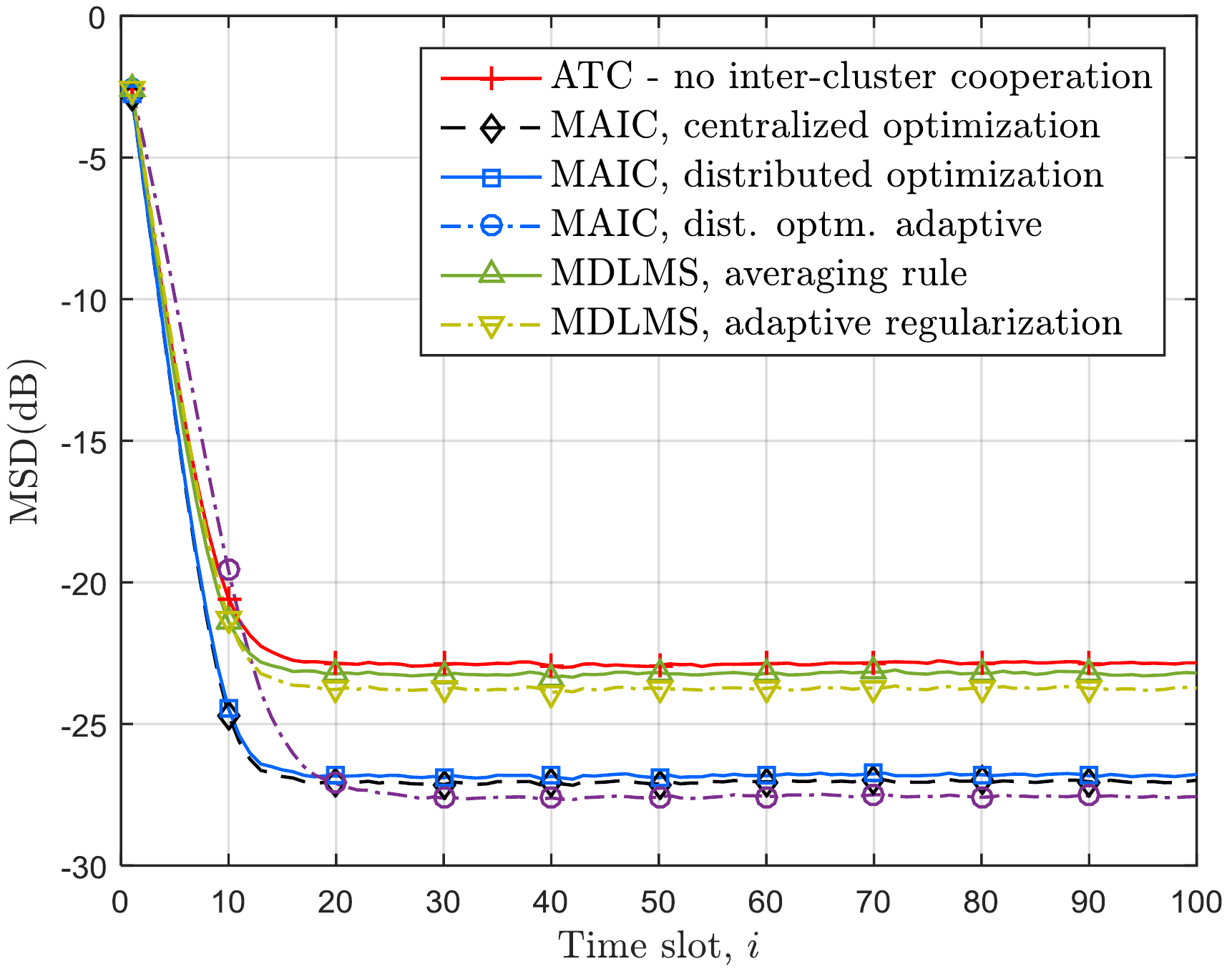}
		\label{fig:cluster_network}}
	\hfil
	\subfloat[Network MSD gains]{\includegraphics[width=3.0in]{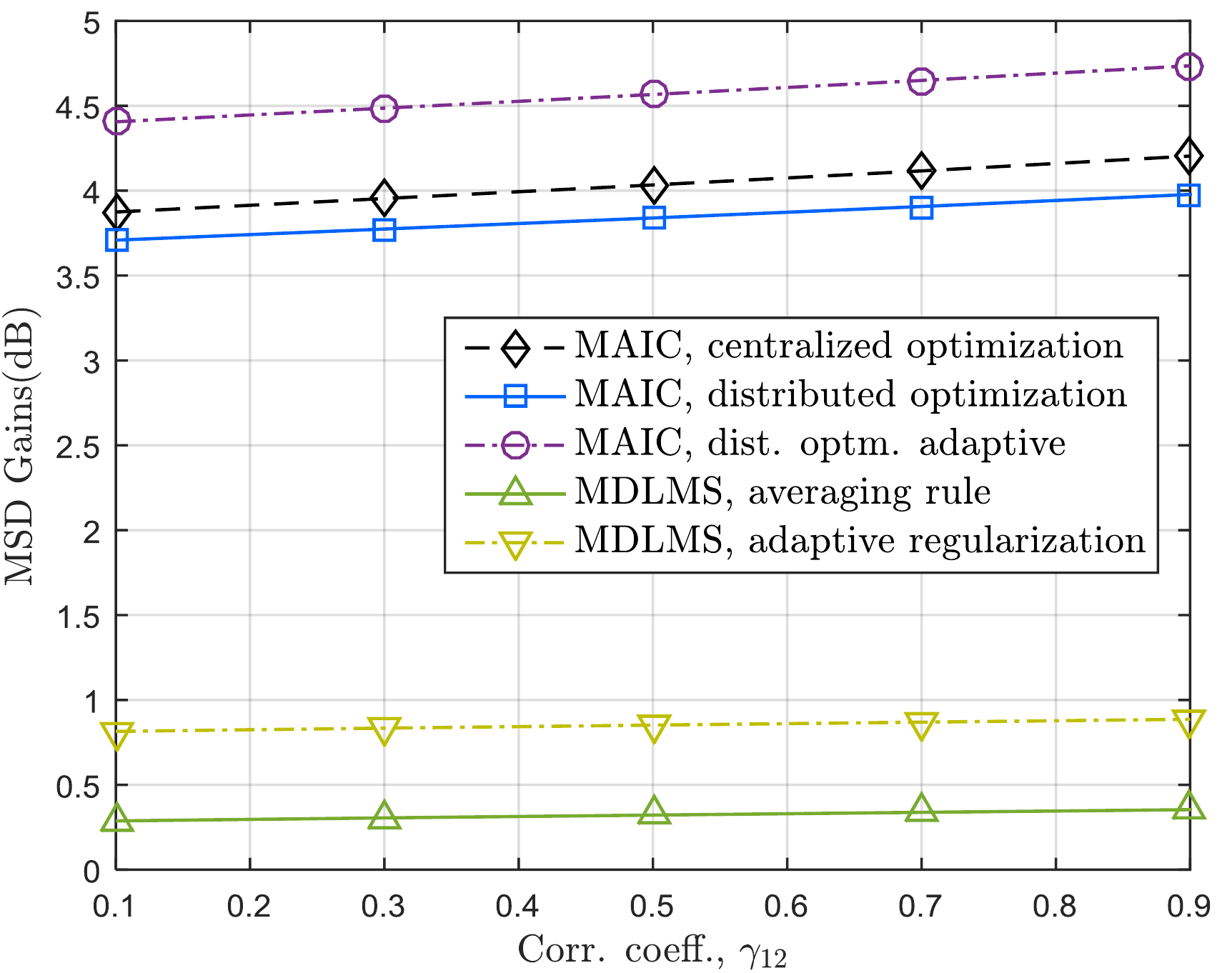}
		\label{fig:msd_gain}}
	\hfil
	\caption{Network MSD (a) and comparison of steady-state network MSD gains with respect to ATC with no inter-cluster cooperation (b) when the regularization weight of MDLMS is $\eta=5$.}
	\label{fig:network_gain}
\end{figure*}
where the correlation coefficients $\gamma_{12}=\gamma_{21}$. The noise variances $\sigma^2_{v,k}$ across all nodes are drawn uniformly from $[-15,\; -5] \text{dB}$. Here, we use $s_v=0.03^2$, step-size $\mu=0.1$, $\eta=5$, and all other conditions remain unchanged as in Section \ref{ssec:simA}. We first fix $\gamma_{12}=0.9$. From Fig.~\ref{fig:cluster}, it can be observed that MDLMS using the averaging rule as well as MDLMS with adaptive regularization lead to a deterioration in the steady-state MSD performance of $C_2$ and $C_3$ compared to the no inter-cluster cooperation ATC, although the overall network performance is improved (see Fig.~\ref{fig:network_gain}(a)). However, the MAIC strategies with optimized weights are able to achieve better steady-state MSDs for all clusters compared to the no inter-cluster cooperation ATC. This clearly shows the benefit of the inter-cluster weights selection scheme proposed in Section \ref{sec:Weights}. Next, we let the correlation coefficient $\gamma_{12}$ range from $0.1$ to $0.9$ to examine how much improvement is achieved against ATC without inter-cluster cooperation by different strategies. As shown in Fig.~\ref{fig:network_gain}, as $\gamma_{12}$ increases, i.e., the parameters $\Bs{w}_{C_1}^\circ$ and $\Bs{w}_{C_2}^\circ$ become more correlated, the MSD gains of the MAIC strategies with optimized weights become larger, whereas the MSD gain of MDLMS with the averaging rule and adaptive regularization do not vary significantly. 

\subsection{Performance Comparison under Different Means}\label{ssec:simC}
\begin{figure*}[!t]
	\centering
	\subfloat[Network MSD under small mean differences]{\includegraphics[width=3.0in]{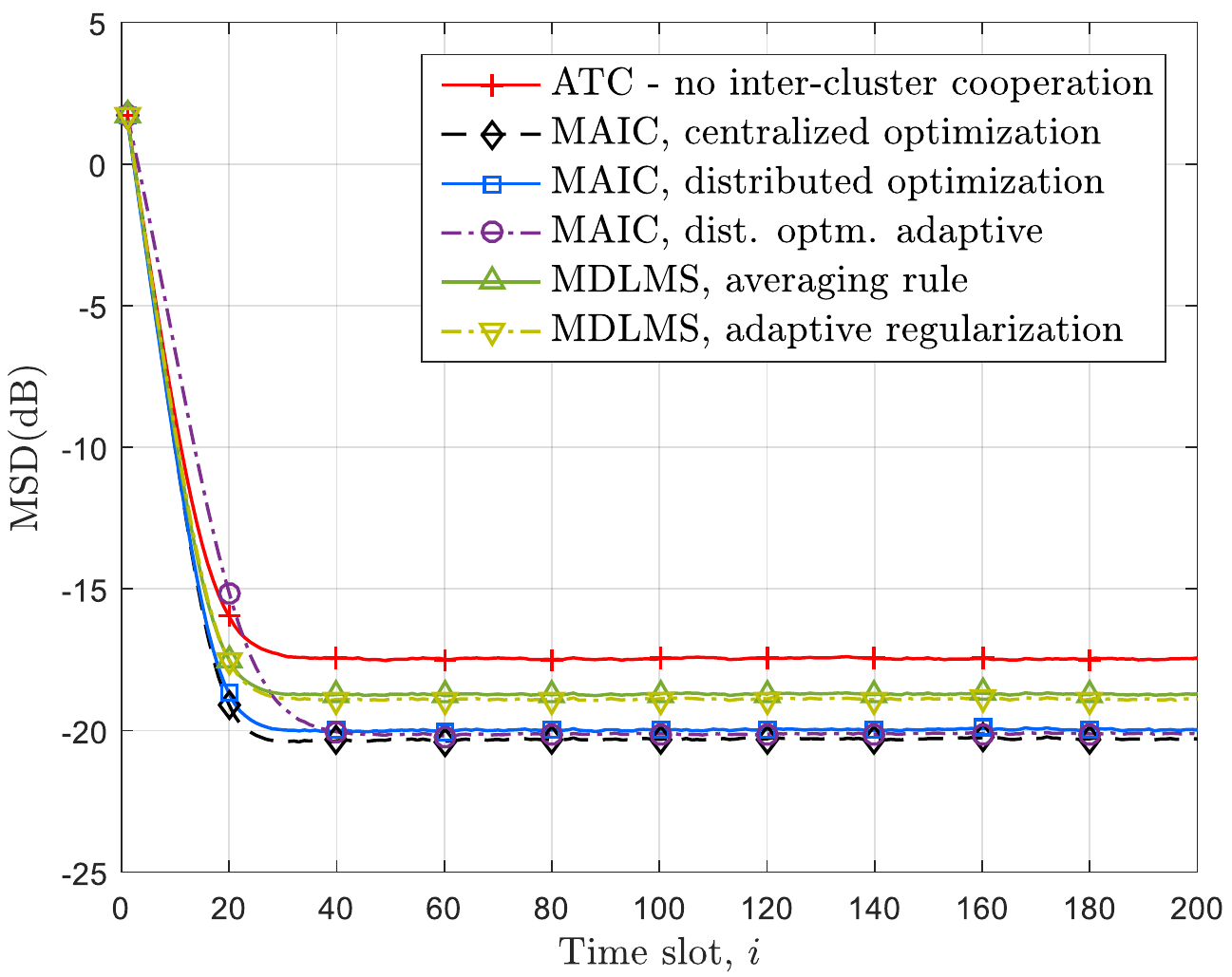}
		\label{fig:small_difference}}
	\hfil
	\subfloat[Network MSD under large mean differences]{\includegraphics[width=3.0in]{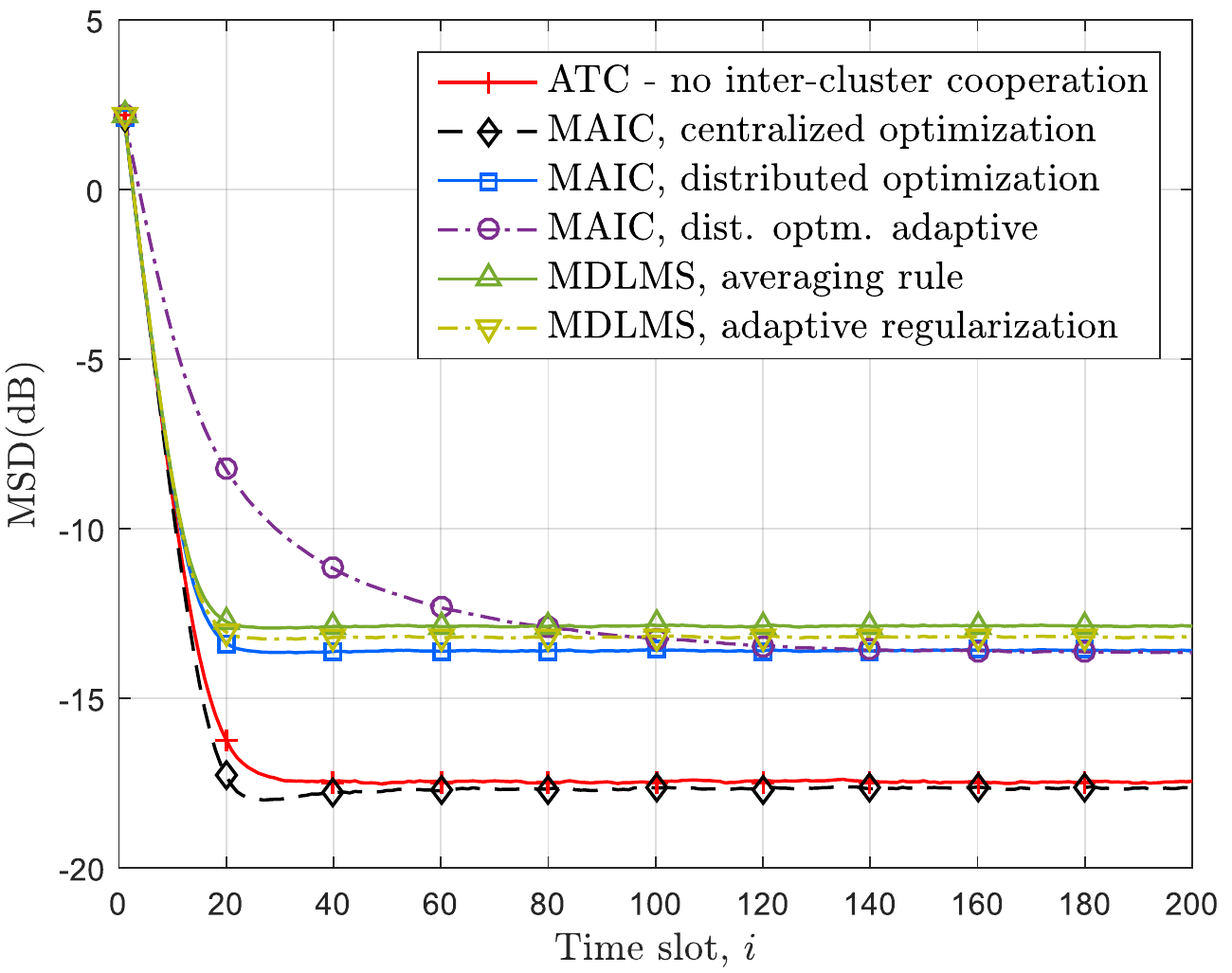}
		\label{fig:large_difference}}
	\hfil
	\caption{Network MSD performance when differences between parameters are small (a) and large (b).}
	\label{fig:different_means}
\end{figure*}
In the next simulation, we compare the performance of each strategy when different cluster parameters have different means. We set $s_v=0.03^2$ and keep all the other settings the same as in Section \ref{ssec:simA}. Choose a $\delta_{\overline{w}} \in (0,1)$ The mean vectors of the three clusters are set to be $\overline{w}_{C_1}=(1-\delta_{\overline{w}})\times\mathds{1}_M$, $\overline{w}_{C_2}=\mathds{1}_M$, and $\overline{w}_{C_3}=(1+\delta_{\overline{w}})\times\mathds{1}_M$. When $\delta_{\overline{w}}=0.06$ is small, it can be observed from Fig.~\ref{fig:different_means}(a) that the MSD of each multitask diffusion strategy is still better than that for ATC without inter-cluster cooperation. When $\delta_{\overline{w}}=0.3$, Fig.~\ref{fig:different_means}(b) shows that the MSDs for all strategies except MAIC with weights obtained by centralized optimization, are worse than the MSD for ATC without inter-cluster cooperation. This is expected since in this case the cluster parameters are on average more different from each other, and inter-cluster cooperation introduces errors into the diffusion process instead of improving the MSD performance. However for MAIC with centralized optimization, the main diagonal entries of $G$ become dominant and the off-diagonal entries tend to zero as $\delta_{\overline{w}}$ becomes large. This allows MAIC to decrease the degree of inter-cluster cooperation to avoid estimation bias so that its performance is similar to that of ATC without inter-cluster cooperation.

\section{Conclusion}\label{sec:Conclusion}
We have proposed a multitask diffusion strategy that performs adaptation before inter-cluster cooperation. We showed by error behavior analysis that this new strategy has the desirable property that mean stability can be achieved regardless of the inter-cluster cooperation weights, which allows these weights to be adjusted without compromising the network stability. We also proposed a centralized as well as distributed method to optimize the inter-cluster cooperation weights in order to improve the steady-state network MSD. Finally, we presented an adaptive implementation, which enables the network to update the inter-cluster cooperation weights according to changes in the noise and data profile as well as the cluster parameter statistics. In this work, we have assumed that node clusters are known a priori. For future research, it would be interesting to investigate the problem of performing adaptive unsupervised clustering simultaneously with the inter-cluster cooperation weight optimization.


%

\appendices
\section{Proof of Theorem \ref{thm:mean}}\label{sec:appdx_A}
Iterating \eqref{eq:meanrecursion} from the time instant $i=1$ and then letting $i\to\infty$ on both sides gives
\begin{align}
\lim\limits_{i\rightarrow\infty} \E \blderr{w}_i = \lim\limits_{i\rightarrow\infty}\CalB^i \;\E\blderr{w}_0 + \lim\limits_{i\rightarrow\infty} \left(\sum_{j=0}^{i-1}\CalB^j\right)r \label{eq:limitmean}
\end{align}
if and only if the spectral radius of the matrix $\CalB$, $\rho(\CalB)$, is less than one. We have
\begin{align}
\Norm{\CalB}_{b,\infty} 
&\leq \Norm{\CalA^T\CalG^T}_{b,\infty} \cdot \Norm{I_{MN} - \mu\CalR_u}_{b,\infty} \nonumber\\
&= \Norm{I_{MN} - \mu\CalR_u}_{b,\infty}, \label{eq:CalBnormInequality}
\end{align}
where the last equality follows since both $A$ and $G$ are left stochastic matrices and $\Norm{\CalA^T\CalG^T}_{b,\infty} = \Norm{A^TG^T}_{\infty} = 1$ from Lemma D.4 of \cite{AHS12TR}. Therefore, if
\begin{align}
\Norm{I_{MN} - \mu\CalR_u}_{b,\infty} < 1, \label{eq:stablecondition}
\end{align} 
then 
$\rho\left( \CalB \right) \leq \Norm{\CalB}_{b,\infty} < 1$, and MAIC achieves mean-stability. The step-size condition \eqref{eq:mean_mu} now follows from \eqref{eq:stablecondition}. 

In addition, when Assumption \ref{asmp:mean} is satisfied, from \eqref{eq:Er} we have $r=\E\Bs{r}=0_{NM}$. From \eqref{eq:limitmean}, we then readily obtain $\lim\limits_{i\rightarrow\infty} \E \blderr{w}_i=0_{NM}$ if $\CalB$ is stable, which implies that MAIC achieves an asymptotically unbiased estimation, and the proof is complete. 

Note that the mean stability condition \eqref{eq:stablecondition} and the resulting condition \eqref{eq:mean_mu} are also valid in the case where the cluster parameters are deterministic. Moreover, if the parameters are deterministic and are the same for every cluster, then it is easy to verify that $r=0$, and both the proposed MAIC and the multi-task diffusion strategy proposed in \cite{JC14TSP} reduce to the conventional single-task ATC strategy where only one cluster (which is the whole network) exists.     

\section{Proof of Theorem \ref{thm:means_quare}}\label{sec:appdx_B}
Iterating the variance relation \eqref{eq:variancerelation} from $i=1$ we have
\begin{align}
\E\Norm{\blderr{w}_{i}}^2_\sigma &= 
\E\Norm{\blderr{w}_{0}}^2_{\CalE^i\sigma} 
+ (f_a+f_b)^T \left(\sum \limits_{j=0}^{i-1} \CalE^j \right)\sigma 
+ \sum_{j=0}^{i-1} \left[f_{c,i-j-1}\right]^T\CalE^j\sigma
{.} \label{eq:variancerelationlimit}
\end{align}
Letting $i\to\infty$, the first and second terms on the R.H.S. of the above equation converges to zero and a finite value, respectively, if and only if $\CalE^i \rightarrow 0$ as $i \rightarrow \infty$, i.e., the matrix $\CalE$ is stable. From \eqref{eq:gc}, we can rewrite the third term of \eqref{eq:variancerelationlimit} as
\begin{align}
2\sum_{j=0}^{i-1} \left[\CalB_I \E\left(\blderr{w}_{i-j-1}\otimes\Bs{r}\right)\right]^T \CalE^j\sigma. \label{eq:msd3rd}
\end{align}
The matrix $\CalB_I$ and vector $\sigma$ in \eqref{eq:msd3rd} have finite entries. In Appendix~\ref{sec:appdx_C}, we show that $\E\left(\blderr{w}_{i-1}\otimes\Bs{r}\right)$ is uniformly bounded if matrix $\CalB$ is stable. In addition, we have
\begin{align}
c_\xi=\Norm{\CalE}_\xi<1,
\end{align}
for some norm $\Norm{\cdot}_\xi$, if matrix $\CalE$ is stable. Therefore, by applying the Cauchy–Schwarz inequality to each term of the sum in \eqref{eq:msd3rd}, and by norm equivalence we have
\begin{align}
\left|\left[\CalB_I \E\left(\blderr{w}_{i-j-1}\otimes\Bs{r}\right)\right]^T \CalE^j\sigma\right|\le a\cdot c_\xi^j, \label{eq:norm_ineq}
\end{align}
for some positive constant $a$. Since $c_\xi^j\rightarrow0$ as $j\rightarrow\infty$, the series,
\begin{align}
\sum_{j=0}^{i-1}\left|\left[\CalB_I \E\left(\blderr{w}_{i-j-1}\otimes\Bs{r}\right)\right]^T \CalE^j\sigma\right|
\end{align}
converges as $i\rightarrow\infty$, which implies the absolute convergence of \eqref{eq:msd3rd}, and the proof is complete.

\section{}\label{sec:appdx_C}
From the error recursion \eqref{eq:errmodel}, applying the Kronecker product with $\Bs{r}$ on both sides we have:
\begin{align}
\blderr{w}_i\otimes\Bs{r}=\big(\BCalB_i\blderr{w}_{i-1}\big)\otimes\Bs{r} - \Bs{s}_i\otimes\Bs{r} + \Bs{r}\otimes\Bs{r} . \label{eq:wor}
\end{align} 
Note that 
\begin{align}
\big(\BCalB_i\blderr{w}_{i-1}\big)\otimes\Bs{r} = \big(\BCalB_{i}\otimes I_{MN}\big)\big(\blderr{w}_{i-1}\otimes\Bs{r}\big),
\end{align}
hence taking expectation on the both sides of \eqref{eq:wor} we obtain
\begin{align}
\E\big(\blderr{w}_i\otimes\Bs{r}\big) = \CalB_I \E\big(\blderr{w}_{i-1}\otimes\Bs{r}\big) + \E\big(\Bs{r}\otimes\Bs{r}\big) {.} \label{eq:Ewor}
\end{align}
Recalling \eqref{eq:r} and \eqref{eq:gb}, and applying the identity $a\otimes a = \Vc(aa^T)$, we find that the last term on the R.H.S.\ of \eqref{eq:Ewor} equals to 
\begin{align}
f_b = \Vc\left[\CalA^T(I_{MN} - \CalG)^T\CalR_w(I_{MN} - \CalG)\CalA \right] {.} \nonumber 
\end{align}
Since $\CalB_I=\CalB\otimes I_{MN}$, thus $\CalB_I$ is stable if matrix $\CalB$ is stable. Therefore, when $\CalB$ is stable, \eqref{eq:Ewor} is a BIBO stable recursion with bounded driving term $f_b$. As a result, $\E\big(\blderr{w}_i\otimes\Bs{r}\big)$ converges and thus is uniformly bounded. Letting $i\rightarrow\infty$, it is easy to obtain from \eqref{eq:Ewor} that
\begin{align}
\E\big(\blderr{w}_\infty\otimes\Bs{r}\big)=\left[(I_{MN}-\CalB)^{-1}\otimes I_{MN}\right]f_b {.} \label{eq:inf_kron_err}
\end{align}

\section{Upper bounds for Steady-state Network MSD used in Section~\ref{ssec:distributed}}\label{sec:appdx_bound}
Since $\CalF$ is required to be stable to ensure the mean-square stability of \eqref{eq:approxvariancerelation}, we have
\begin{align}
\left(I_{M^2N^2}-\CalF \right)^{-1} = \sum_{j=0}^{\infty} \CalF^j, \label{eq:Fsum}
\end{align}
Then substituting \eqref{eq:matrixF} and \eqref{eq:Fsum} into \eqref{eq:MSDapprox}, and applying identities $\Vc(AXB)=(B^T\otimes A)\Vc(X)$ and $\Tr(AB) = \Vc(A^T)^T\Vc(B)$, we have 
\begin{align}
\hat{\zeta} = \frac{1}{N}\sum_{j=0}^{\infty} \Tr \left({\CalB}^j \CalX {\CalB^T}^j\right), \label{eq:upperbound_cent_app1}
\end{align}
where matrix $\CalX$ is given by
\begin{align}
\CalX &= \mu^2\CalA^T\CalG^T\CalS\CalG\CalA + \CalA^T(I_{MN} - \CalG)^T\CalR_w(I_{MN} - \CalG)\CalA .
\end{align}
From Lemma D.~2 of \cite{AHS12TR}, we obtain 
\begin{align}
\frac{1}{N}\sum_{j=0}^{\infty} \Tr \left({\CalB}^j \CalX {\CalB^T}^j\right)
&\le \frac{b}{N}\cdot\sum_{j=0}^{\infty}\Norm{\CalB^j}_{b,\infty}\cdot\Norm{\CalX}_{b,\infty}\Norm{{\CalB^T}^j}_{b,\infty} \nonumber\\
&\le  b\cdot \left(\sum_{j=0}^{\infty} \Norm{\CalB}_{b,\infty}^{2j} \Norm{\CalX}_{b,\infty} \right)\nonumber\\
&\le  b\cdot \left(\sum_{j=0}^{\infty} \beta^{2j} \Norm{\CalX}_{b,\infty} \right)\nonumber\\ 
&\le  \frac{b}{1-\beta^2}\cdot \Norm{\CalX}_{b,\infty} \label{eq:upperbound_cent_2nd}\\
&\le  c\cdot \Tr\left( \CalX \right) ,  \label{eq:upperbound_cent_app2}
\end{align}
where we use \eqref{eq:CalBnormInequality} in the third inequality, $b$ and $c$ are positive constants, and $\beta=\Norm{I_{MN}-\mu\CalR_u}_{b,\infty}<1$. The R.H.S. of \eqref{eq:upperbound_cent_app2} is the bound \eqref{eq:upperbound_cent} in Section \ref{ssec:distributed}. 

Next, we assume that the matrix $A$ is symmetric so that $\CalA$ is as well. From the \eqref{eq:upperbound_cent_2nd}, we obtain
\begin{align}
\frac{1}{N}\sum_{j=0}^{\infty} \Tr \left({\CalB}^j \CalX {\CalB^T}^j\right) 
&\le \frac{b}{1-\beta^2}\cdot \Norm{\CalX}_{b,\infty} \nonumber\\
&\le \frac{b}{1-\beta^2}\cdot \Norm{\CalA^T}_{b,\infty}^2 \cdot \Norm{\CalY}_{b,\infty} \nonumber\\
&= \frac{b}{1-\beta^2}\cdot \Norm{\CalY}_{b,\infty} \nonumber\\
&\le \bar{c}\cdot \Tr\left( \CalY \right) ,  \label{eq:upperbound_cent_app3}
\end{align}
where $\bar{c}$ is a positive constant. In the third inequality we use the fact $\Norm{\CalA^T}_{b,\infty}=1$, and the matrix $\CalY$ is given by
\begin{align}
\CalY = \mu^2 \CalG^T\CalS\CalG + (I_{MN} - \CalG)^T \CalR_w (I_{MN} - \CalG).
\end{align}
This is the bound \eqref{eq:upperbound_dist} in Section \ref{ssec:distributed}.




\bibliographystyle{IEEEtran}

\end{document}